\documentstyle[12pt]{article}
\begin{document}
\baselineskip=22.pt plus 0.2pt minus 0.2pt
\lineskip=22.pt plus 0.2pt minus 0.2pt

\begin{titlepage}

\renewcommand{\thefootnote}{\fnsymbol{footnote}}
\setcounter{footnote}{0}

\begin{flushright}
CU-TP-598, LBL-32682
\end{flushright}
\begin{center}
{\bf\LARGE Multiple Collisions and Induced Gluon Bremsstrahlung
in QCD \footnote{
This work was supported by the Director, Office of Energy Research,
Division of Nuclear Physics of the Office of High Energy and
Nuclear Physics of the U.S. Department of Energy under Contract Nos.
DE-AC03-76SF00098, DE-FG05-90ER40592, and DE-FG02-93ER40764.}
}\\[5ex]

{\large Miklos Gyulassy} \\
{\em Physics Department, Columbia University, New York, NY 10027\\
and\\}
{\large Xin-nian Wang}\\
{\em Nuclear Science Division, MS 70A-3307, Lawrence Berkeley Laboratory\\
University of California, Berkeley, CA 94720}\\[2ex]
\today\\[2ex]
\end{center}

\begin{abstract}
\baselineskip=20.pt plus 0.2pt minus 0.2pt

Induced soft gluon bremsstrahlung associated with
multiple collisions is calculated via perturbative
QCD. We derive the non-abelian analog of the Landau-Pomeranchuk effect
that  suppresses induced soft  radiation
with formation times exceeding the mean free path, $\lambda$.
The dependence of the  suppression effect
on the $SU(N)$ representation of the jet parton as well as
the kinematic variables is expressed through a
radiation formation factor. Unlike in QED, the finite contribution
from the small $x$ regime in QCD  leads to
an approximately constant  radiative energy loss
per unit length, $dE/dz\propto \mu^2$, in the high energy
limit that is sensitive to the
infrared screening scale, $\mu$, of the medium.
As a function of the dimensionless parameter
$\zeta=\lambda \mu^2/E$, we show furthermore how
the energy dependence of
$dE/dz$ evolves from the above  constant  for $\zeta\ll 1$
to the more familiar (Bethe-Heitler) linear dependence for $\zeta \gg 1$.
\end{abstract}

\end{titlepage}

\baselineskip=22.pt plus 0.2pt minus 0.2pt

\section{Introduction}

Radiative energy loss of  ultra-relativistic particles
passing through dense matter is of  interest not only because of its
many practical applications, but also
because it illustrates a characteristic  destructive
interference phenomenon
caused by  the finite formation time\cite{lpm,akhiezer},
\begin{equation}\tau(k)\sim \hbar/\Delta E(k)
\sim 2\omega/k_\perp^2 \sim 2/\omega\theta^2
\; \; ,\label{tauk} \end{equation}
of quanta with large four momentum, $k^\mu=(\omega,k_z,{\bf k}_\perp)$,
emitted at small angles, $\theta\approx k_\perp/\omega$,
relative to the incident particle.
In effect, $\tau(k)$, is the minimal time needed
to resolve the transverse wavepacket
of the quanta with, $\Delta x_\perp\sim \hbar/k_\perp$,
from the wavepacket
of its  high energy ($E_0\gg \omega)$ parent.
Destructive interference between radiation amplitudes
associated with multiple collisions can  be expected when the
 mean free path, $\lambda$, is short compared to the formation time.
When $\tau(k)\gg \lambda$, the emitted quanta cannot
resolve different elastic scattering centers, and
the assumption of independent contributions from each separate
scattering in the medium breaks down.
This effect, first studied in  QED and then in other
field theories, is often referred to
as the Landau-Pomeronchuck-Migdal (LPM) effect\cite{lpm,akhiezer}.

Interest in analogous destructive interference
phenomena in non-abelian theories
is connected with  attempts to understand the weak nuclear dependence
of hard  QCD  processes
and the use of those dependencies as a probe of
the space-time development of hadronization.
A high energy quark or gluon
passing through  dense (QCD)  matter of course
suffers multiple interactions. However,
in the $Q\rightarrow \infty$ limit,
QCD factorization theorems\cite{mueller} apply that show that,
as in the QED case, the soft  radiation is emitted  only from the external
legs\cite{marchesini}. There are a large number of phenomena
such as the very weak nuclear dependence of Drell-Yan
yields\cite{bodwin,moss}
and the apparent nuclear independence
of very high energy quark fragmentation\cite{eA,plumer},
that confirms this basic factorization feature of {\em asymptotic} pQCD.
As shown in  Ref.\cite{bodwin},  however,
the assumptions leading to  the factorization theorem
break down for sub-asymptotic conditions. In the Drell-Yan process,
for example, final state interaction corrections become
important for sufficiently large nuclei with,
$A^{2/3}\Lambda_{QCD}^2> M^2_{DY}$.
In deep inelastic hadroporduction, nuclear dependence
always exits in the $x_F<0$ region and disappears in
the $x_F>0$ region only when $\kappa R/\nu \rightarrow 0$,
where $\kappa\approx 1$ GeV/fm\cite{eA,plumer}.

In order to understand  such nuclear effects more quantitatively,
 it is necessary to study induced radiation patterns associated with
multiple collisions in QCD.
Rough estimates for the magnitude of the suppression
of non-abelian induced radiation  have been made in
\cite{highpt,sorensen,gavin,ryskin,brod2}.  The
estimates  for $dE/dz$
vary, however,  widely ranging from  energy independent
\cite{brod2},
logarithmic energy dependent\cite{highpt}
to $dE/dz\propto E$ \cite{gavin,ryskin}.
However, a detailed study of the LPM effect in QCD has not been
performed to our knowledge
in the context of multiple collision theory,
taking into account essential non-abelian features of the problem.
The aim of this paper is to initiate such a study.

While  pQCD can serve only as a qualitative guide because the effective
coupling,
$\alpha_s$, is  small only in extremely dense matter, e.g., a quark-gluon
plasma at temperatures $T\gg T_c \sim
\Lambda_{QCD}$,  it is instructive
 to explore its consequences in situations where approximations
to multiple scattering can be used to simplify the problem.
Additional  motivation
for this work is to compare radiation patterns due to multiple collision
physics in  pQCD with those suggested
by phenomenological string models\cite{plumer,lund,hijing}
for
high energy $eA, pA,$ and $AA$ reactions.
We show, for example, that some features of the  induced soft
gluon rapidity distributions in pQCD
are  similar to the hadron
distributions predicted  by multi-string Lund type models utilizing
string breaking and flip mechanisms\cite{plumer}.
In particular the effective string tension, $\kappa\sim 1$ GeV/fm
in those models is analogous to
constant radiative energy loss due to induced  soft radiation
in the $x\ll 1 $ regime  in non-abelian multiple
collision theory. Furthermore, additivity of radiation from
multiple scatterings is limited to a domain $x< \kappa \lambda/E_0$,
that shrinks as $E_0$ increases, just as in string flip models
for multiple interactions\cite{plumer}.
However, in pQCD, unlike in string models,
 $\kappa= dE/dz \propto \mu^2 \sim g^2 T^2$,
is found to be sensitive to the infrared screening scale,
$\mu$, in the medium.
Recall that other  so called  ``string
effects'' were also found to arise naturally
from interfering pQCD amplitudes  for three jet events in $e^+e^-$
\cite{mueller}. Finally, we note  the importance of clarifying
radiative energy loss mechanisms and interference phenomena associated
with formation zone physics for the further development of
QCD transport theory and parton cascade models\cite{transport}.

Before turning to the non-abelian problem, we
recall briefly how the destructive interference for induced radiation
occurs in QED. The Fourier transform
of the current of a classical particle
undergoing $m$ collisions at space-time points $x_i^\mu$
at which the four momenta change from $p_{i-1}^\mu$ to $p_{i}^\mu$
is
\begin{equation}J^\mu(k)=\sum_{i=1}^m J^\mu_i(k)=
ie\sum_{i=1}^m e^{ikx_i}\left(\frac{p_i^\mu}{kp_i}
- \frac{p_{i-1}^\mu}{kp_{i-1}}\right)
\; \; . \label{qedj} \end{equation}
The resulting  soft bremsstrahlung spectrum is given by
\begin{equation}\omega\frac{d^3 n_\gamma}{d^3 k}=
\sum_{\epsilon} |\epsilon J(k)|^2 /(2(2\pi)^3)=-|J(k)|^2/(2(2\pi)^3)
\; \; . \label{qeddn} \end{equation}
The sum contains  $m$ diagonal terms, where the phase factors drop out,
and $m(m-1)$ off diagonal terms involving phase factors,
$\exp(ik(x_i-x_j))$. Two extreme limits are obvious.
One is the incoherent limit where $k(x_i-x_j)\gg 1$.
In this case,
the off diagonal phase factors tend to average to zero. This corresponds
to the usual  Bethe-Heitler limit, in which
the radiated energy loss,
$dE/dz=-E/L_r$, grows linearly with energy
with $L_r$ being the radiation length
(the ratio of the mean free path to the average fraction
of the  energy radiated per collision).
The other extreme limit, is the one corresponding to
 $k(x_i-x_j)\ll 1$ for all $i,j$.
In this case,  all the phase factors
are approximately unity, and there is an exact cancellation
between adjacent terms in eq.
(\ref{qedj}). Only the radiation
from initial and final lines contributes. This is
the so-called the ``Factorization limit''
 since the amplitude for soft radiation
factors into an amplitude for multiple collisions
times a current element depending only on the momenta of
the external lines entering and leaving the reaction.
Because the radiation intensity in the high $Q^2=-(p_f-p_i)^2$
limit  increases only as $\log(Q^2)$, the radiated energy
loss grows only logarithmically with the number, $m$,
of elastic collisions
in the medium. Therefore, for a random walk leading to
$Q^2\propto m$, the induced radiation energy loss
per collision, $dE/dm \propto \log(m)/m$, becomes negligible
for large $m$.
 In the general case, between these extreme limits,
there is a partial contribution
from intermediate current elements.

The essential parameter controlling this interference effect
is the ratio of the formation time (\ref{tauk})
to the distance between multiple interactions.
For an ultra-relativistic particle
propagating in a straight line
\begin{equation}k(x_i-x_j)=(\omega-k_z)L_{ij}=L_{ij}/\tau(k)
\; \; , \label{qed2} \end{equation}
where $L_{ij}=z_i-z_j$ is the longitudinal distance (time)
between scattering at $i$ and $j$. For finite deflection
angles, an additional phase, $-{\bf k}_\perp\cdot{\bf r}_{ij}$,
appears that depends on the transverse separation of the scattering events.
Interference between
current elements
$J_i$ and $J_j$ occurs only if $k(x_i-x_j)\ll 1$. This requires
$\tau(k) \gg L_{ij}$ and  ${k}_\perp\ll {r}_{ij}$. The
interference pattern depends also on the current
correlation function, $\langle J_i(k)J_j(-k) \rangle$.
The LPM effect in QED often  refers to
the  specific destructive interference pattern
calculated by Migdal using  the
Fokker-Planck transport equation to
solve for  the probability distribution, $W({\bf x},{\bf p},
{\bf p}^\prime)$, of scattering points
and initial and final momenta.
 Monte Carlo methods have also been developed \cite{stanev} to calculate
the development of very high energy  cosmic ray air showers.
The interference effect found in the limit that
the scattering medium is much thicker than the radiation length
is that the familiar soft $1/\omega$ bremsstrahlung frequency spectrum
is transformed into  a $1/\surd \omega$ form for $\omega < E_0^2/E_{LPM}$.
Remarkably, because the  characteristic energy,
$E_{LPM} \sim 3\;{\rm TeV} \sim 5\;{\rm ergs}$,
turns out to be so large,
this interesting prediction has yet to be verified quantitatively
experimentally\cite{spencer}.

In the following sections, we calculate  the induced non-abelian radiation
for a high energy parton passing through
the random color field produced by a color neutral ensemble
of static partons. This idealized system is chosen
to minimize the complications of multiple scattering theory
while illustrating  the  essential features of
of the non-abelian LPM effect.
In section 2, we first calculate the  elastic multiple  collision
amplitude for a spinless high energy jet parton belonging to an arbitrary
$SU(N)$ representation.
 We  show how classical multiple
collision cascade theory emerges from  pQCD in both the high and low momentum
transfer regions in the limit where  the mean free path is large compared
to the range of the Debye screened potentials.
In section 3, we calculate the soft gluon radiation amplitudes
in the restricted kinematic region range $ k_\perp \ll \mu$.
This restriction limits the applicability of the results
to ``thin'' plasmas, with the number of mean free paths
$m$ not large, or to very small $x\ll 1/\sqrt{m}$ gluons.
The extension of the results to thick plasmas, that would
be necessary to make contact
with the Migdal ($R=\infty,E_0\rightarrow \infty$) limit\cite{lpm},
is not considered in this paper.
The soft eikonal approximations used  here are aimed, on the other hand,
to the study of the breakdown of factorization
 in
the opposite limit ($E_0=\infty,R\rightarrow \infty)$.
For high energy reactions involving finite nuclei, this
is in fact the only physically relevant limit for applications
of pQCD in any case.
Destructive interference between the radiation amplitudes
from jet and exchanged gluon lines is shown to limit
the transverse momentum distributions, and
the non-abelian generalization of (\ref{qedj}) is  derived.
In section 4, we introduce and calculate the ``radiation formation factor''
controlling the magnitude of the suppression of induced  radiation in pQCD.
We emphasize the role of color algebra on the destructive interference
pattern. In section 5, the radiative energy loss, $dE/dz$, due to soft
induced radiation   is estimated. Finally, a discussion of remaining
open problems is presented in section 6.

\section{Multiple Elastic Scattering in a Color Neutral Ensemble}

\subsection{The Model Potential}

Consider
the sequential elastic scattering of a high energy (jet)
parton  in the random color field produced by an ensemble of $m$
static partons located at ${\bf x}_i=(z_i,{\bf x}_{\perp i})$
such that $z_{i+1}>z_i$ and $(z_{i+1}-z_i)\gg \mu^{-1}$,
where
$\mu$ is the color screening mass in the medium.
As a simplified
model of multiple scattering in  a color neutral quark-gluon plasma, we
 assume a  static Debye screened potential for each target parton:
\begin{equation}
V_i^{a}({\bf q})= g(T^a_i)_{c,c^\prime} e^{(-i{\bf q}\cdot{\bf x}_i)}
/({\bf q}^2
+\mu^2)
\; \; , \label{aia} \end{equation}
where $T^a_i$
is a $d_i$-dimensional generator of $SU(N)$
corresponding to the representation of the
target parton  at ${\bf x}_i$.
The initial and final color indices,
$c, c^\prime$, refer to the target  parton are  averaged
and summed over when computing  the ensemble averaged cross sections.
With $V_i^{a}\propto T^a_i$ the  ensemble averaged potential
vanishes everywhere,  $\langle V_i^a \rangle\propto TrT^a_i= 0$.
However, since
\begin{equation}Tr{T}^a_i{T}^b_j=\delta_{ab}\delta_{ij}
({d}_i/d_A){C}_{2i}
\; \; . \label{tai} \end{equation}
the diagonal mean square fluctuations and the cross sections are finite.
Recall that for $SU(N)$  the second order Casimir, ${C}_{2i}=(N^2-1)/2N\equiv
C_F$
for quarks in the fundamental ($d_i=N)$ representation,
while ${C}_{2i}=N\equiv C_A$ for gluons in the adjoint ($d_i=N^2-1\equiv d_A$)
representation.

In this potential,  each scattering leads on the average to only
a relatively small momentum transfer $q_i^\mu=(q_i^0,q_{zi}, {\bf q}_{\perp i})
$ with each component being much less than
the incident energy, $E_0$.
The assumption that the  potentials are  static is approximately
valid in a high temperature  plasma
of  massless quarks and gluons in the following sense:
As  $T\rightarrow \infty$, the effective coupling $g\rightarrow 0$
(albeit very slowly). The perturbative
 Debye screening mass $\mu\sim gT$ limits
$q_\perp \stackrel{<}{\sim} gT$. The typical thermal energy
$E_T\sim 3T$ of the plasma constituents is therefore large
compared to $\mu$. Consequently, the average energy loss per
elastic collision,
$ -q^0 \approx -q^z \approx q_\perp^2/2 E_T \propto
g^2 T $, is $\sim g$ times smaller than the average transverse
momentum transfer.

Because we are interested in relatively low momentum transfer scattering
($\Lambda_{QCD}\ll q_\perp\sim gT \ll T)$,
 the  spin
of the partons can be neglected.
 The jet parton is allowed, however,
 to be in an arbitrary $d$-dimensional representation of $SU(N)$
with generators, $T^a$, satisfying  $T^aT^a=C_2{\bf 1}_d$.

The  Born (color matrix) amplitude  to scatter from an incident
four momentum $p_{i-1}^\mu$
to $p_i^\mu$ in the potential centered at ${\bf x}_i$ is then given by
\begin{equation}M_i(p_i,p_{i-1})=2\pi\delta(p^0_i-p^0_{i-1})A_i({\bf q}_i)
e^{-i{\bf q}_i\cdot{\bf x}_i}
\; \; , \label{vi} \end{equation}
where ${\bf q}_i={\bf p}_i-{\bf p}_{i-1}$, and $A_i$ is shorthand for
\begin{equation}A_i({\bf q}_i)= T^a A_i^a ({\bf q}_i)=-2igE_0 T^a V_i^{a}({\bf
q}_i)
\; \; . \label{ai} \end{equation}
The differential cross section 	averaged
over initial and summed over final colors of both projectile
and target partons reduces to the familiar form for
low transverse momentum transfers:
\begin{equation}d\sigma_{i}/d q_{\perp i}^2 \approx
 C_i\frac{4\pi\alpha^2}{(q_{\perp i}^2+\mu^2)^2}
\; \; , \label{dsigi} \end{equation}
 where the color factor is
\begin{equation}C_i=\frac{1}{d d_i}
Tr({T}^a{T}^b)Tr (T^a_i{T}^b_i)= C_2 {C}_{2i}/d_A
\; \; . \label{ci} \end{equation}
For $SU(3)$, $2C_i$ gives the usual
color factors
$4/9,1,9/4$ for $qq,qg,gg$ scattering respectively.
In our notation, the angular distribution is given by
\begin{equation}d\sigma_{i}/d\Omega_i=\frac{1}{d d_i}Tr
|A_i({\bf q}_i)|^2/(4\pi)^2
\; \; . \label{dsido} \end{equation}

\subsection{Sequential Multiple Scattering}

Our main assumption for computing
the multiple elastic scattering amplitude
is that the scattering centers are  well
separated in the sense $L_i=z_{i+1}-z_i \gg \mu^{-1}$.
In a  quark-gluon plasma
at very high temperature $T$, $1/\mu\sim 1/gT$ and
the effective $qg$ scattering  cross section  from eq.(\ref{dsigi})
is  $\sigma \sim 2 \pi\alpha^2/\mu^2
\sim g^2/8\pi T^2$.  Given a Stefan-Boltzmann
 density of partons $\rho\sim 5T^3$,
the mean free path is $\lambda \sim 5/(g^2 T)\gg d\sim 1/gT$
for $g\ll  1$. Hence, $L_i \gg d$ is satisfied at extreme
temperatures at least.

The dominant Born amplitude for coincident {\em sequential}
scattering with target partons from $i$ to  $j$
without radiation is  then simply
\begin{equation}M_{ji}(p_j,p_{i-1}) = \int \frac{d^4 p_i}{(2\pi)^4} \cdots
\frac{d^4 p_{j-1}}{(2\pi)^4}
M_j(p_j,p_{j-1})i\Delta(p_{j-1})\cdots i\Delta(p_i)M_i(p_i,p_{i-1})
\; \; , \label{mij} \end{equation}
where $\Delta(p)=(p^2-m^2+i\epsilon)^{-1}$ is the intermediate
jet parton propagator. Amplitudes involving  backscattering are suppressed at
high energies because of the limited momentum transfer that
each potential can impart.
Because of the energy delta function in eq.(\ref{vi}),
the integrations over the intermediate $p_i^0$ set
all of them to $E_0$ and lead to a
 conservation factor
\begin{equation}\delta(ji)\equiv 2\pi\delta(p_j^0-p_{i-1}^0)
\; \; . \label{delij} \end{equation}
Therefore,
\begin{eqnarray}
M_{ji}(p_j,p_{i-1}) &=&\delta(ji) \int \frac{d^3{\bf p}_i}{(2\pi)^3} \cdots
\frac{d^3{\bf p}_{j-1}}{(2\pi)^3}e^{-i({\bf p}_j-{\bf p}_{j-1})\cdot{\bf x}_j}
A_j({\bf p}_j-{\bf p}_{j-1})  \nonumber \\
&\;& \hspace{1in} \times \left\{ \prod_{k=i}^{j-1}
e^{i\pi/2} e^{-i({\bf p}_k-{\bf p}_{k-1})\cdot{\bf x}_k} \frac{A_k({\bf
p}_k-{\bf p}_{k-1})}{
P_0^2-{\bf p}_k^2+i\epsilon} \right\}
\; \; , \label{mij2} \end{eqnarray}
where $P_0=(E_0^2-m^2)^{1/2}\approx E_0$, and the product is path ordered
from left to right with decreasing index $k$.
Rearranging the phases in terms of the separation vectors
\begin{equation}
{\bf R}_k\equiv L_k\hat{e}_z + {\bf r}_k ={\bf x}_{k+1} -{\bf x}_{k}
\; \; , \label{vr}\end{equation}
 we can write
\begin{equation}M_{ji}(p_j,p_{i-1}) =\delta(ji) e^{-i{\bf p}_j\cdot{\bf x}_j}
e^{+i{\bf p}_{i-1}\cdot{\bf x}_i}I_{ji}(p_j,p_{i-1})
\; \; , \label{defi} \end{equation}
where  the reduced amplitude is
\begin{equation}I_{ji}(p_j,p_{i-1})=
\int \left\{ \prod_{k=i}^{j-1}
\frac{d^3{\bf p}_k}{(2\pi)^3}
 \frac{e^{i\pi/2}e^{+i{\bf p}_k \cdot{\bf R}_k}}{
P_0^2-{\bf p}_k^2+i\epsilon}\right\}
A_j({\bf p}_j-{\bf p}_{j-1})  \cdots A_i({\bf p}_i-{\bf p}_{i-1})
\; \; . \label{mij3} \end{equation}
Because  of the assumed ordering, $L_k>0$, and the  integrals
over $p_{zk}$ can be evaluated by closing the contour in the upper half
plane, setting the intermediate jet legs on-shell with
\begin{equation}p_{zk}=(P_0^2-p_{\perp k}^2)^{1/2}
\approx P_0-p_{\perp k}^2/2P_0
\; \; . \label{pzk} \end{equation}
The singularities of the $A_k$ at $p_{zk}\approx P_0+i(q_{\perp k}^2
+\mu^2)^{1/2}$ can be neglected because they leave very
small residues $\propto \exp(-\mu L_k)$ given the assumed
large separation $\mu L_k\gg 1$ between scattering centers.
Therefore,
 \begin{equation}I_{ji}(p_j,p_{i-1})\approx
\int \left\{ \prod_{k=i}^{j-1}
\frac{d^2{\bf p}_{\perp k}}{(2\pi)^2}
 \frac{e^{+ i{\bf p}_{k}\cdot {\bf R}_{k}} }{2P_0}
\right\}
A_j({\bf p}_j-{\bf p}_{j-1})  \cdots A_i({\bf p}_i-{\bf p}_{i-1})
\; \; , \label{mij40} \end{equation}
with $p_{zk}$ given by eq.(\ref{pzk}). Note that in the high energy limit
$I_{ji}$ survives in spite of the $1/P_0$ residues because
$A_k\propto E_0 $ due to the vector nature of the coupling.
Also the ordering of the potentials in eq.(\ref{mij2},\ref{mij40})
in decreasing order of the index cannot be permuted  in the non-abelian
case because
of the non-commuting color matrices in the $A_i$.

{}From eq.(\ref{mij40}) we can derive two interesting
limits.  One is
 the semi-classical (large angle) cascade  limit, and
the other is the eikonal (straight line) limit for  multiple
small momentum transfer scatterings.
 The first case illustrates how the transverse
momentum integrations can decouple resulting in a factorized
form of the multiple collision amplitude and is discussed in Appendix
A. The second limit is however physically more relevant and
is considered below .

\subsection{Color Algebra}

To make explicit the color algebra we write
\begin{equation}M_{ji} = (a_j\cdots a_i)M_{ji}^{a_j\cdots a_i}
\; \; , \label{mijcol} \end{equation}
in a   shorthand notation where
\begin{equation}(a_j\cdots a_i)\equiv T^{a_j}\cdots T^{a_i}
\; \; , \label{brack} \end{equation}
and we adopt the usual summation convention over repeated indices.
For the color neutral ensemble under consideration
\begin{equation}M_{ji}^{a_j \cdots a_i}\propto
(T^{a_j}_j)_{c_jc_j^\prime}\cdots (T^{a_i}_i)_{c_ic_i^\prime}
\; \; . \label{colors} \end{equation}
Hence,
\begin{equation}
\langle M_{ji}^{a_j \cdots a_i}
(M_{ji}^{a^\prime_j \cdots a^\prime_i})^\dagger
\rangle \propto \prod_{k=i}^j \frac{1}{d_k}Tr(T^{a_k}_k T^{a_k^\prime}_k)
\propto \prod_{k=i}^j \left(\delta_{a_k a_k^\prime}
 C_{2k}/d_A \right)
\; \; . \label{ensembl} \end{equation}
Given eq.(\ref{ensembl}),
the color factor associated with the jet parton
is given by
\begin{equation}C(i,j)=\frac{1}{d} Tr((a_i\cdots a_j)(a_j\cdots a_i)) =
C_2^{j-i+1}
\; \; , \label{cij} \end{equation}
as obtained by repeated use of the basic $(aa)=C_2{\bf 1}_d$ relation.
This is simply the product of the color factors, $C_2$, occurring
 for each isolated
collision as in eq.(\ref{ci}).
Therefore, even though the amplitude
eq.(\ref{mijcol}) does not factor in color space, the ensemble
averaged coincidence cross section does factor
for large angle elastic scattering in a locally
color neutral ensemble as shown in appendix A.

\subsection{Eikonal Limit}

For a high energy jet, the coincidence amplitude is
dominated by small angle scattering.
In  this case,  we
change variables to
${\bf q}_l ={\bf p}_l-{\bf p}_{l-1}$
with  $p_{zl}\approx P_0\approx E_0$ approximately fixed, and
write the total momentum transfer as
\begin{equation}{\bf Q}_{ji} \equiv \sum_{l=i}^{j} {\bf q}_l={\bf p}_j-{\bf
p}_{i-1}
\; \; . \label{endq} \end{equation}
The coincidence amplitude
can then reduces to
 \begin{eqnarray}
M_{ji}(p_j,p_{i-1})
&\approx& \delta(ji) (a_j\cdots a_{i})(-ig)^{j-i+1}2E_0
\int \left\{ \prod_{k=i}^{j}
\frac{d^2{\bf q}_{\perp k}}{(2\pi)^2}
 e^{-i{\bf q}_k \cdot {\bf x}_{k}}
 V_k^{a_k}({\bf q}_k) \right\} \nonumber \\
&\;& \hspace{1in} \times
(2\pi)^2 \delta^2({\bf Q}_{\perp ji}-\sum_{l=i}^{j} {\bf q}_{\perp l})
\; \; . \label{mij41} \end{eqnarray}
Note that the  dependence of the phase on the $z_k$
can be factored out with
\begin{equation}\sum_{k=i}^j q_{zk} z_k \approx (p_{zj}-P_0)z_j +
(P_0 -p_{z,i-1})z_i
\; \; . \label{phasz} \end{equation}
This phase is important only for off-shell amplitudes with
$p_{zj}\ne P_0$ or $p_{z,i-1}\ne P_0$.

To average over the transverse coordinates ${\bf x}_k$,
we employ the
frozen target approximation
taking the initial and final wavefunction of target
parton, $k$, to be $\phi_{ki}({\bf x}_k)$ and
by $\phi_{kf}({\bf x}_k)$ respectively. The amplitude to leave
the target in
a specific final state is obtained by replacing the phase factors
by transition form
factors
\begin{equation}e^{-i{\bf q}_k\cdot{\bf x}_k} \rightarrow F^k_{if}({\bf
q}_k)=\int d^3x_k \phi^*_{kf}({\bf x}_k)
e^{-i{\bf q}_k\cdot{\bf x}_k} \phi_{ki}({\bf x}_k)
\; \; . \label{form} \end{equation}
 After squaring $M_{ji}$, we must sum
over all final states $\phi_{kf}$. For scattering in a chaotic
or thermal bath
we must also average over an ensemble of initial $\phi_{ki}$.

A simplification is possible   in the high energy limit
when the energy and longitudinal momentum transfers
are small, and they can be neglected or replaced
 by their average values
 in $\delta(ji)$ and $F^k_{if}$. In that case, closure
( $\sum_f \phi_{kf}\phi_{kf}^*=1$) can be applied to
the  sum over final states. Averaging in addition
 over initial states with probability
$p(i)$, the average squared amplitude contains factors such as
\begin{equation}\sum_i p(i) \sum_f F^k_{if}({\bf q}_{\perp k}) F^k_{fi}(-{\bf
q}_{\perp k}^\prime)
=\int d^3 x_k
e^{-i({\bf q}_{\perp k}-{\bf q}_{\perp k}^\prime ) \cdot{\bf x}_k}
\rho_k({\bf x}_k) \equiv T_k(
{\bf q}_{\perp k}-{\bf q}_{\perp k}^\prime )
\; \; , \label{close} \end{equation}
where $\rho_k({\bf x}_k)=\sum_i p(i) |\phi_{k0}({\bf x}_k)|^2$ is the ensemble
average density distribution of target parton $k$.
In eq.(\ref{close}) note that $T_k({\bf q}_{\perp})$ is just
the Fourier transform of the Glauber thickness function
\begin{equation}T_k({\bf x}_{\perp k}) = \int dz \; \rho_k(z,{\bf x}_{\perp k})
\; \; , \label{tkg} \end{equation}
which is the probability per unit area of finding parton  $k$
at a transverse coordinate ${\bf x}_{\perp k}$.
For a broad $z$ distribution, it would appear that we may have violated
the assumed $\mu L\gg 1$ assumption. However, in the $m!$ different $z$
orderings of
 the centers, one of the previously neglected
 backscattering amplitudes
becomes dominant and after relabelling the dummy indices
the same result is recovered. The only
essential assumption is that  the mean free path
is long compared to the range of the potential.

With eq.(\ref{close}), the ensemble average of the squared amplitude
is proportional to
 \begin{eqnarray}
\langle |M_{ji}(p_{i-1},p_j)|^2 \rangle
&\propto &
\int \left\{ \prod_{k=i}^{j}
\frac{d^2{\bf q}_{\perp k}}{(2\pi)^2}\frac{d^2{\bf q}_{\perp
k}^\prime}{(2\pi)^2}
 T_k({\bf q}_{\perp k}-{\bf q}_{\perp k}^\prime)
 V_k^{a_k}({\bf q}_k)(V_k^{a_k}({\bf q}_k^\prime))^* \right\} \nonumber \\
&\;& \hspace{0.2in} \times
(2\pi)^4 \delta^2({\bf Q}_{\perp ji}-\sum_{l=i}^{j} {\bf q}_{\perp l})
\delta^2({\bf Q}_{\perp ji}-\sum_{l=i}^{j} {\bf q}_{\perp l}^\prime)
\; \; . \label{mij410} \end{eqnarray}
Only diagonal color components survive because of the color neutrality
condition eq.(\ref{ensembl}).
If the transverse coordinates are distributed over a radius,
$R\gg \mu^{-1}$, then the Fourier transform of the thickness function
will limit the difference,
 $|{\bf q}_{\perp k} -{\bf q}_{\perp k}^\prime|
{\mathrel{\lower.9ex\hbox{$\stackrel{\displaystyle <}{\sim}$}}}
1/R $. Because $V_k^{a_k}({\bf q}_k)$ varies slowly on a scale
$1/R \ll \mu$, we can therefore approximate
$(V_k^{a_k}({\bf q}_k^\prime))^*\approx (V_k^{a_k}({\bf q}_k))^*$
in the integrand. The ${\bf q}_k^\prime$ integrals result
therefore in a multiplicative geometrical factor
\begin{equation}\int \left\{ \prod_{k=i}^{j}
\frac{d^2{\bf q}_{\perp k}^\prime}{(2\pi)^2}
 T_k({\bf q}_{\perp k}-{\bf q}_{\perp k}^\prime) \right\}
(2\pi)^2\delta^2(\sum_{l=i}^{j} ({\bf q}_{\perp l} - {\bf q}_{\perp
l}^\prime) )=
\int d^2 {\bf b} \prod_{k=i}^{j}T_k({\bf b})
\; \; . \label{mij42} \end{equation}
With the above simplification,
the ensemble averaged coincidence  cross section
to scatter sequentially with partons from $i$ to $j$ reduces  to
\begin{equation}d\sigma_{ji}/d^2{\bf Q}_{\perp ji}=
\int d^2{\bf b} \int \prod_{k=i}^{j}\left\{
d^2{\bf q}_{\perp k} T_k({\bf b}) d\sigma_{k}/d^2{\bf q}_{\perp k} \right\}
\delta^2({\bf Q}_{\perp ji}-\sum_{l=i}^{j} {\bf q}_{\perp l})
\; \; . \label{eikonal} \end{equation}
This is recognized as
 the classical Glauber multiple collision limit, with
$p_k({\bf b})=\int T_k({\bf b}) d\sigma_{k}$
being the probability of scattering off center $k$ on a classical trajectory
at impact parameter ${\bf b}$.

\section{Induced Soft Non-Abelian Radiation}

We turn  next to the inelastic amplitudes for induced
radiation of a gluon with color $c$ and light cone momenta and polarization
\begin{eqnarray}
k^\mu &=&(\omega,k_z,{\bf k}_\perp)=[xP^+,k_\perp^2/xP^+,
{\bf k}_\perp] \nonumber \\
\epsilon^\mu
&=&(\epsilon_0,-\epsilon_0,\vec{\epsilon}_\perp)=[0,2\vec{\epsilon}_\perp
\cdot {\bf k}_\perp/xP^+,\vec{\epsilon}_\perp]
\; \; . \label{kinem} \end{eqnarray}
Light cone coordinates are denoted here by square brackets,
$[k^+,k^-,{\bf k}_\perp]$,
with $k^\pm = \omega \pm k_z=k_\perp^2/k^\mp$.
We chose the two physical polarization states for on shell
$(k^2=0)$ gluons to satisfy both
$\epsilon k=0$ and $\epsilon n=0$ with $n^\mu=[0,2,0_\perp]$
in terms of two orthonormal $\vec{\epsilon}_\perp$. Thus
$\epsilon^0=\vec{\epsilon}_\perp
\cdot {\bf k}_\perp/(\omega+k_z)$.
In light cone coordinates, the incident jet parton has
$p_0^\mu=[P^+,m^2/P^+,0]$
with $P^+\approx 2E_0$.
We focus  on the soft limit defined by
 $x\ll 1$. First
we consider the amplitudes
for radiation from the high energy parton lines.
Then  we show that the three gluon amplitudes
essentially cut off the soft $dk_\perp/k_\perp$
spectrum at $k_\perp
\sim  \mu$.
Note that the induced bremsstrahlung associated with a single
isolated collision was derived in pQCD in Ref.\cite{gunion}.
Our interest here is on the induced radiation pattern associated with multiple
collisions.

\subsection{Radiation from Internal Jet lines}

The amplitude to emit a gluon with color $c$ from the $j$th intermediate
jet line during sequential scattering with target partons
from 1 to $m$ is
\begin{eqnarray}
M_{m1}^{cj}(k,p_m,p_0) &=& \int \frac{d^4p_j}{(2\pi)^4}
M_{m,j+1}(p_m,p_j-k)   \nonumber \\
&\; & \times \left\{
i \Delta(p_j-k)
(-2ig\epsilon p_j T^c)
 i \Delta(p_j)
 \right\} M_{j,1}(p_j,p_0)
\; \; . \label{mrim0} \end{eqnarray}
The integrals
over $p_j^0$ set their values to $E_0$ because of the $\delta(1j)$
in $M_{j,1}$.
However, for all the subsequent intermediate lines in $M_{m,j+1}$
the energy is shifted from $E_0$ to $E_0-\omega$, and
a new overall energy conservation factor arises
\begin{equation}\tilde{\delta}(m1) \propto \delta(p_m^0 - E_0 +\omega)
\; \; . \label{delm} \end{equation}
These shifted energies change the classical momenta in subsequent
legs $(k>j)$ to ${{\bf P}}_k= {P}_\omega \hat{\bf R}_k$ with
\begin{eqnarray}
{P}_\omega &=&((E_0-\omega)^2 -m^2)^{1/2}\approx P_0 - \omega/v_0
\; \; . \label{pkm} \end{eqnarray}
where
$v_0=P_0/E_0$ is the speed of the incident parton.

To perform the $p_{zj}$ integral, it is convenient
to split the two propagators
using
\begin{equation}2\epsilon p_j
\Delta(p_j-k)\Delta(p_j)= \frac{\epsilon (p_j-k)}{
k(p_j-k)}\Delta(p_j-k)- \frac{\epsilon p_j}{
k_\mu p_j^\mu}\Delta(p_j)
\; \; ,\label{del} \end{equation}
which is valid  for on shell radiation since  $k^2=0$ and $\epsilon k=0$.
The contour integral over $p_{zj}$ can then
be performed as discussed in Appendix B.

With (\ref{tilij1},\ref{tilij2}) from Appendix B,
the radiation amplitude eq.(\ref{mrim0}) reduces in the kinematic region
$x\ll 1$ and $k_\perp \ll \mu$ to
\begin{eqnarray}
M^{cj}_{m1}  &\approx& g\tilde{\delta}(m1) e^{-i\omega t_m}
e^{-i{\bf p}_m\cdot{\bf x}_m}e^{+i{\bf p}_0\cdot {\bf x}_1}
\int \frac{d^2{\bf p}_{\perp j}}{(2\pi)^2}
\frac{1}{2P_0}
e^{i({\bf P}_j\cdot{\bf R}_j-\frac{L_j}{2P_j}{\bf p}_{\perp j}^2) }\nonumber \\
&\; & \hspace{0.2in} \times
\left(
\frac{\epsilon \tilde{p}_j}{
k\tilde{p}_j} e^{+ikx_j} -
\frac{\epsilon p_j}{k
p_j}  e^{+ikx_{j+1}} \right)
I_{m,j+1}(p_m,p_{j}) T^c I_{j,1}(p_j,p_0)
\; \; . \label{mrip3} \end{eqnarray}
Note the appearance of the phases
$kx_j$ and $kx_{j+1}$ where $x_j^\mu=(t_j,{\bf x}_j)$ and  the interaction
time, $t_j$, is the classical transit time
along the path from ${\bf x}_1$ to ${\bf x}_j$
as defined in eq.(\ref{tk}).
In addition, the current element in the brackets involves
on-shell momenta, $p^\mu_j,\tilde{p}^\mu_j$, defined by
\begin{equation}p_j^\mu=(E_0,(P_0^2-({\bf P}_{\perp j}
+{\bf p}_{\perp j})^2)^{1/2},
{\bf P}_{\perp j}+{\bf p}_{\perp j})
\; \; , \label{pjmu1} \end{equation}
\begin{equation}\tilde{p}_j^\mu=(E_0-\omega,
({P}_\omega^2-({\bf P}_{\perp j}+{\bf p}_{\perp j})^2)^{1/2},
{\bf P}_{\perp j}+{\bf p}_{\perp j})
\; \; . \label{pjmu2} \end{equation}
Note finally the order in which
 the color matrix $T^c$ appears above.

For radiation with  $k_\perp\ll \mu$,
we can  factor out of the integrand a current element
proportional to  $\epsilon P_j/kP_j$. To see this, note
first that for
a high energy on-shell parton with
$p=[(1-\delta)P^+,m_\perp^2/(1-\delta)P^+,{\bf p}_\perp]$
and radiation with kinematics (\ref{kinem}),
\begin{equation}\frac{\epsilon p}{kp}=2\frac{\vec{\epsilon}_\perp
\cdot ({\bf k}_\perp-x{\bf p}_\perp/(1-\delta))}{
({\bf k}_\perp-x{\bf p}_\perp/(1-\delta))^2+ x^2m^2/(1-\delta)^2}
\approx 2\frac{\vec{\epsilon}_\perp
\cdot {\bf k}_\perp}{k_\perp^2}      \; \;
\; \;  {\rm for}\; \;x{m}_\perp \ll
k_\perp
\; \; . \label{app} \end{equation}
This approximate independence of the current element
on $p^\mu$  allows us to factor out (\ref{app}) from
the integrals ${\rm for}\; \;x{m}_\perp \ll
k_\perp$.
However, a more general expression can be
factored out that is valid also in the  high
momentum transfer limit. For fixed ${\bf x}_i$ and $E_0\rightarrow \infty$,
 the
momenta $p_j^\mu$ and $\tilde{p}_j^\mu$ are approximately fixed by geometry
to be $ P_j^\mu\approx E_0(1,\hat{\bf R}_j)$, and therefore
\begin{equation}\frac{\epsilon p_j}{kp_j}\approx \frac{\epsilon
\tilde{p}_j}{k\tilde{p}_j} \approx \frac{\epsilon P_j}{kP_j}
\; \; . \label{curre} \end{equation}
Since this expression also reduces to (\ref{app}) when
$\hat{\bf R}_j$ points along the jet direction,
the final factorized form of the amplitude
for intermediate line radiation becomes
\begin{eqnarray}
M_{m1}^{cj}&\approx& g\frac{\epsilon P_j}{kP_j}\left(e^{ik x_j}
-e^{ikx_{j+1}}\right) (a_m\cdots a_{j+1}ca_j\cdots a_1) M_{m1}^{a_m\cdots a_1}
\; \; . \label{mrf} \end{eqnarray}
In this expression, the $j$ independent
phase factor, $\exp({-i\omega t_m})$, was discarded.

We note several points in connection  the amplitudes
for induced radiation from the internal jet lines given
by eq.(\ref{mrf}).
\begin{enumerate}
\item{ The approximate factored expression
holds both in the large angle cascade
limit considered in Appendix A  and straight line
(Eikonal) limits of  $M_{m1}$}
as long as  $x\ll 1$ and $k_\perp\ll \mu$.
\item{ Just as in the QED case, these intermediate
line amplitudes vanish for radiation with formation length
significantly exceeding the separation of adjacent
scattering centers. In particular,
for fixed $x=k^+/P^+\approx \omega/E_0 \ll 1$ the phase
factors cancel  in the $k_\perp\rightarrow 0$ limit:
\begin{equation}M_{m1}^{cj}
\propto k(x_{j+1}-x_j)=\omega L_j(1/v_0-\cos \theta)-
{\bf k}_\perp\cdot{\bf r}_{\perp j} \approx
L_j/\tau(k) - {\bf k}_\perp\cdot{\bf r}_{\perp j}\rightarrow 0
\; \; , \label{limr} \end{equation}
where $\tau(k)$ is the formation time (length)
from eq.(\ref{tauk}).}
\item{ Unlike in momentum space, there is no factorization
in color space. The color matrix for the amplitude
without radiation is  modified
by radiating a gluon of color $c$ after the $j$th interaction
by the insertion of a $T^c$ matrix:
\begin{equation}(a_m\cdots a_j\cdots a_1)
\rightarrow (a_m\cdots c a_j\cdots a_1)
\; \; . \label{modc} \end{equation}}
\item{ Even for  $x\ll 1$ the condition leading to the approximate
form in (\ref{app}) breaks down for very large number of collisions
since the random walk in transverse momentum
space leads to a growing $\langle p_{\perp m}^2 \rangle \propto  m\mu^2$.
Thus for large $m$ the above approximation is  only valid
in a restricted  $x$ region
\begin{equation}
x \ll \frac{1}{\surd{m}} \frac{k_\perp}{\mu} \ll \frac{1}{\surd{m}}
\; \; . \label{shrink}\end{equation}}
\item{  We must also
add the amplitudes for radiation from the initial
and final lines  to the above amplitudes for radiation
from internal lines. These external line amplitudes are
 \begin{eqnarray}
M_{m1}^{c0}&=& M_{m1}(p_m,p_0-k)(i\Delta(p_0-k))(-2ig\epsilon p_0T^c)
\nonumber \\
&\approx& -g\frac{\epsilon P_0}{kP_0} e^{ik x_1}
 (a_m\cdots a_1 c) M_{m1}^{a_m\cdots a_1} \nonumber \\
M_{m1}^{cm}&=& (i\Delta(p_m))(-2ig\epsilon p_mT^c)M_{m1}(p_m,p_0)
\nonumber \\
&\approx& g\frac{\epsilon P_m}{kP_m} e^{ik x_m}
 (c a_m\cdots a_1 ) M_{m1}^{a_m\cdots a_1}
\; \; , \label{mrf2} \end{eqnarray}
where we defined $P_0^\mu=p_0^\mu,P_m^\mu=p_m^\mu$ for notational
convenience,
and we again discarded the  common phase factor, $\exp({-i\omega t_m})$,
as in eq.(\ref{mrf}).}
\end{enumerate}

The sum of all the amplitudes (\ref{mrf},\ref{mrf2}) can
then be written in a suggestive form
\begin{equation}M_m^c(k,p_m,p_0)\approx -i[\epsilon J(k)]^c_{a_m\cdots a_1}
M^{a_m\cdots a_1}_{m1}(p_m,p_0)
\; \; , \label{sum} \end{equation}
where the effective color current element analogous to (\ref{qedj}) is
\begin{equation}
[\epsilon J(k)]^c_{a_m\cdots a_1}=ig\sum_{l=1}^m e^{ikx_l}
\left( \frac{\epsilon P_l}{k P_l} (a_m\cdots c a_l \cdots a_1)
-\frac{\epsilon P_{l-1}}{k P_{l-1}} (a_m\cdots a_l c \cdots a_1)\right)
\; \; , \label{curr} \end{equation}
In the abelian case, the matrices in $(\cdots)$ are set to unity,
and Eqs.(\ref{sum},\ref{curr})
reduce to the  soft radiation formulas of QED.
In the abelian case the current actually
vanishes as a power of $x$ in the $x\rightarrow 0$ limit
on account of (\ref{app}).
In the non-abelian case,
the non-commutativity of the generators
leads, however, to a non-vanishing current in (\ref{curr}) even for
 $x=0$.

We emphasize that the approximate form of the effective
current in  eq.(\ref{curr}) is
valid only in a restricted kinematic domain (\ref{shrink}).
Gauge invariance requires the
absence of induced radiation associated
with collisions without momentum transfer\cite{gunion},
i.e., for $P_l=P_{l-1}$. On the other hand, the contribution
from scattering at $l$ in eq.(\ref{curr}) is non-vanishing
if $[c,a]\ne 0$.
To recover full gauge invariance, of course all the amplitudes involving
three and four gluon
vertices as well must be added to the above result.
However, since  the domain
of applicability of eq.(\ref{curr}) shrinks to zero
as $ q_{\perp j}\rightarrow 0$, it is consistent with
the gauge invariance requirement in the kinematic domain indicated.

For the physically most interesting eikonal limit,
 the  effective color current
reduces for $x \surd{m}\ll
k_\perp/\mu
\ll 1  $ to
\begin{eqnarray}
[\epsilon J(k)]^c_{a_m\cdots a_1}&\approx& 2ig\frac{\vec{\epsilon}_\perp
\cdot {\bf k}_\perp}{k_\perp^2} \sum_{l=1}^m e^{ikx_l}
(a_m\cdots [c,a_l]\cdots a_1)
\; \; . \label{curr2} \end{eqnarray}
The lower bound on the domain of applicability comes from
(\ref{shrink}).
For $k_\perp< x\mu\surd{j}$, the effective current has components
in the directions ${\bf p}_{\perp j}$ that cannot be factorized out
of the elastic amplitude.
We note that (\ref{curr2}) can also be derived directly from (\ref{mrim0})
using the eikonal form (\ref{mij41}) for  $M_{m,j+1}$
and $M_{j,1}$. The above derivation has the advantage
that the connection to the familiar abelian case is made more transparent.

\subsection{Radiation from Internal Gluon Lines}

In the previous section
we concentrated on the   amplitudes for induced
radiation  in which the gluon is radiated directly from
the fast jet lines without final state interactions.
Here we consider  amplitudes involving one or more  three or four
gluon vertices in which the gluon scatters with one or more
of the target partons before emerging with
 kinematics given by (\ref{kinem}).
In the high energy limit, the  amplitudes involving the four gluon vertices
can be generally neglected because those vertices are momentum independent
and a contact interaction with two widely separated target partons  is small.
The amplitudes  involving three gluon vertices
can be classified by the number of such vertices and the indices of the
scattering centers to which one of the gluon legs is  attached.
The simplest  of such amplitudes, denoted by $G_{jm}^c$,
corresponds to a (possibly virtual) gluon  emitted by the jet
between the $z_{j-1}$ and $z_{j+1}$
with that gluon   scattering off the target parton at $z_j$.
 For $1<j<m$,
\begin{eqnarray}
G_{mj}^c&=&\int \frac{dp_{j}^4}{(2\pi)^4}\frac{dp_{j-1}^4}{(2\pi)^4}
M_{m,j+1}(p_m,p_j)i\Delta(p_j) G^c_{j}(k,p_j,p_{j-1})
i\Delta(p_{j-1})
M_{j-1,1}(p_{j-1},p_0) \nonumber \\
 &\;& \hspace{2in}
\; \; , \label{gjm} \end{eqnarray}
where with $q_j=p_j-p_{j-1}$ the single
three gluon amplitude in the Feynman gauge is
\begin{eqnarray}
G^c_{j}(k,p_j,p_{j-1})&=&( -igT_{b}(p_{j-1}+p_j)^{\alpha})
(-i\Delta(q_j))
\nonumber \\
&\; & \hspace{0.2in} \times(-gf_{ba_jc}
\Lambda_{\alpha\beta\gamma}(q_j,-q_j-k,k))
 A^{a_j\beta}_{j}(q_j+k)) \epsilon^\gamma(k)
\; \; . \label{mcj} \end{eqnarray}
The external field at $j$ in our case is
\begin{equation}A^{a_j\beta}_{j}(q)= g^{\beta 0} 2\pi\delta(q^0)
V^{a_j}_j({\bf q}) e^{-i{\bf q}\cdot{\bf x}_j}
\; \; , \label{externj} \end{equation}
and the three gluon tensor  is
\begin{equation}\Lambda^{\alpha\beta\gamma}(p_1,p_2,p_3)=(p_2-p_3)^\alpha
g^{\beta\gamma}
+ (p_3-p_1)^\beta g^{\gamma\alpha} + (p_1-p_2)^\gamma g^{\alpha\beta}
\; \; . \label{lam3} \end{equation}

The  amplitudes with multiple three gluon vertices
 correspond to  multiple final state interactions
of the emitted gluon.
Because the centers are assumed to be far apart,
the intermediate gluon lines in those amplitudes are  set on shell
(by the corresponding $dp_z$ contour integral).
Those amplitudes therefore describe
final state cascading of the gluon in the medium.
In QED these type of amplitudes are replaced by
higher order
Compton like amplitudes.
For gluons, final state cascading
will broaden their final
$k_\perp$ distribution and induce further gluon showering.
However, for the problem of the
 energy
loss of the incident jet, which is our primary interest here,
the final transverse momentum distribution of the rescattered
gluons is not important. In addition, in the soft limit
$k_\perp \ll \mu$ each triple gluon vertex
gives rise to a factor $O(k_\perp/\mu)$ smaller
than the corresponding internal jet line radiation
amplitude, as we  show below.
We therefore concentrate here only  on
the amplitudes involving one three gluon vertex given
by (\ref{gjm}).

As these amplitudes only arise in the non-abelian case,
we can simplify the derivation by
evaluating (\ref{gjm}) in the eikonal limit.
As shown in Appendix C, we find in this limit that
\begin{eqnarray}
G^c_{jm} &\approx& 2\pi \delta(p_m^0-E_0+\omega)
(-ig)^{m} (2E_0)
 \int \prod_{k=1}^m \left(
\frac{d^2 q_{\perp k}}{(2\pi)^2}
 e^{-i{\bf q}_{\perp k}\cdot {\bf x}_{\perp k}}
V^{a_k}_k({\bf q}_{\perp k}) \right) \nonumber \\
 &\;&\times(2\pi)^2
\delta({\bf Q}_{\perp } - {\bf k}_{\perp} - \sum_{k=1}^m {\bf q}_{\perp k})
 \left\{ 2ge^{iz_j/\tau(k)}\frac{\vec{\epsilon}_\perp\cdot
({\bf q}_{\perp j}-{\bf k}_\perp)}{({\bf q}_{\perp j}-{\bf k}_\perp)^2}
(a_m\cdots[c,a_j]\cdots a_{1}) \right\} \nonumber \\
&\; & \hspace{2in} . \label{eik4} \end{eqnarray}
We note that corrections to the effective current element
in the $\{\}$ brackets
arise   for  large $j$ in the region $k_\perp < x p_{\perp j}
\sim x \mu \surd{j}$ from terms neglected in the vertex function
in eq.(\ref{gamma2}) of Appendix C.
In eq.(\ref{eik4}), ${\bf Q}_\perp= {\bf p}_{\perp m}$
is the final transverse momentum of the jet parton.

Summing these  three gluon amplitudes and adding
the radiation amplitudes from the jet lines in
Eqs.(\ref{sum},\ref{curr2}), we  obtain
the total amplitude for $m$-fold coincidence scatterings together
with soft radiation  in the eikonal limit:
\begin{eqnarray}
M^c_m(k,p_m,p_0)&\approx&
2\pi \delta(p_m^0-E_0+\omega)
(-ig)^{m} (2E_0)
\int \prod_{k=1}^m \left(
\frac{d^2 q_{\perp k}}{(2\pi)^2}  e^{-i{\bf q}_{\perp k}\cdot {\bf x}_{\perp
k}}
V^{a_k}_k({\bf q}_{\perp k}) \right) \nonumber \\
 &\;&\; \; \; \; \times(2\pi)^2
\delta({\bf Q}_{\perp } - {\bf k}_{\perp} - \sum_{k=1}^m {\bf q}_{\perp k})
\; \left\{ -i\vec{\epsilon}_\perp \cdot \vec{J}^c_{a_1\cdots
a_m}(k;\{{\bf q}_{\perp i}\})  \right\} \nonumber \\
\; \; \; , \label{sum2} \end{eqnarray}
where the effective color current is
\begin{equation}\vec{J}^c_{a_1\cdots a_m}(k;\{{\bf q}_{\perp i}\})=
\sum_{i=1}^m e^{iz_i/\tau(k)} \vec{\jmath}_i(k) (a_m\cdots [c,a_i]
\cdots a_1)
\; \; , \label{curr3} \end{equation}
and the elementary current elements are
\begin{equation}\vec{\jmath}_i(k)= 2ig\left( \frac{{\bf k}_\perp}{k_\perp^2}+
\frac{{\bf q}_{\perp i}-{\bf k}_{\perp}}{|{\bf q}_{\perp i}-{\bf k}_\perp|^2}
 \right)
\; \; . \label{jelem} \end{equation}
The following points should be noted in connection with the above results:
\begin{enumerate}
\item{For the case $m=1$,
eq.(\ref{curr3})  reduces to the result derived  in \cite{gunion}.}
\item{For $k_\perp \ll q_{\perp i}$,  the
three gluon amplitudes can be neglected
as noted before. Therefore,
 in this limit eq.(\ref{curr3}) reduces to eq.(\ref{curr2}).}
\item{However, for very small  $k_\perp< x\mu_{\perp i}$
corrections to eq.(\ref{jelem}) arise as can be seen from eq.(\ref{app}).
In particular, the  singularities at $k_\perp=0 $
is  regulated on a scale
$x\mu$, where  $\mu$ is the (dynamic) mass of the jet parton.}
\item{The singularity at $k_\perp = q_{\perp i}$
is a non-abelian
feature  due to  induced radiation along the direction
of  the exchanged gluon.
It is regulated by the
 gluon polarization tensor in the medium\cite{selik3}.}
\item{The approximate color current is strictly
 valid only for $k_\perp \ll \mu$ and $x\ll 1$.
However, eq.(\ref{jelem}) shows the general cancellation
of amplitudes for $k_\perp \gg q_{\perp i}$
that  limits the  induced radiation  from an isolated scattering
to $k_\perp \stackrel{<}{\sim} \mu $. }
\item{For  $q_{\perp i}=0$, the current element $\vec{\jmath}_i$ vanishes
in accordance with  gauge invariance\cite{gunion}.}
\item{The phase factor in eq.(\ref{curr3}) is independent of the
transverse coordinates ${\bf x}_{\perp i}$ in the eikonal limit.
The  transverse phase factors,
$\exp(-i{\bf k}_{\perp}\cdot{\bf x}_{\perp i})$, associated
with each isolated collision    are spread over  the net elastic phase
factor $\exp(-i\sum_k {\bf q}_{\perp k}\cdot{\bf x}_{\perp k})$.}
\item{Finally,  corrections to eq.(\ref{jelem}) also appear
 in   kinematic domains
where either the radiation formation time, $\tau(k)$, or the
intermediate jet lifetime, $2\omega/({\bf q}_{\perp i}
-{\bf k}_\perp)^2$,
is on the order of the mean free path
due as shown by eq.(\ref{sumphas}) in Appendix C}.

\end{enumerate}

Eqs.(\ref{curr3},\ref{jelem}) are the main result
of this paper  from which we derive next  the non-abelian
LPM interference effect  in
the eikonal limit.

\section{The Radiation Formation Factor}

The spectrum of soft induced
bremsstrahlung associated with multiple scattering in a color neutral
ensemble eq.(\ref{ensembl})
can be computed from eq.(\ref{curr3}) using steps similar
to those leading from eq.(\ref{mij41}) to eq.(\ref{eikonal}).
Analogous to eq.(\ref{qeddn}), we find that
\begin{eqnarray}
\omega \frac{d^3 n_m}{d^3 k} &=& \frac{1}{2(2\pi)^3}
\frac{1}{C_2^m d}
 \left\langle Tr \sum_{\vec{\epsilon}_\perp}\left|[\epsilon J(k,\{
{\bf q}_{\perp i}\})]^c_{a_1\cdots a_m}\right|^2
\right\rangle
\; \; . \label{rad} \end{eqnarray}
Recall that $C_2^m$ is the color factor for the coincidence
scattering cross section without radiation from eq.(\ref{cij})
with  $C_2$ and $d$ being  the second order Casimir and dimension
of the $SU(N)$ representation of the jet parton.
As in eq.(\ref{mij410}) the assumptions of color neutrality
and that the transverse distribution of target partons
is much wider than $\mu^{-1}$ are essential to obtain the above
diagonal form in color and ${\bf q}_{\perp i}$ labels.
Note that the squared current involves a sum over repeated
color indices, $c,a_i$, and the trace is over the resulting sum
of products of color matrices that we consider in detail below.
The average, denoted by large brackets, above is over
 the transverse momentum transfers, ${\bf q}_{\perp i}$,
and given in our case by
\begin{equation}
\langle f({\bf q}_{\perp i}) \rangle
=\int \left\{ \prod_{i=1}^m \frac{\mu^2 d^2{\bf q}_{\perp i}}{
\pi(q_{\perp i}^2 +\mu^2)^2}
 \right\} f({\bf q}_{\perp i})
\; \; . \label{eikave} \end{equation}
We are mainly  interested in comparing the induced
spectrum for $m>1$ to the radiation spectrum
from a single isolated collision\cite{gunion}:
\begin{equation}
\omega \frac{d^3 n_1}{d^3 k}=
\left\langle\frac{\alpha_s C_A}{\pi^2}
\frac{q_\perp^2}{k_\perp^2({\bf k}_\perp -{\bf q}_\perp)^2}\right\rangle_+
\; \; . \label{gunion1} \end{equation}
The average in this case is over a single scattering
as obtained from eq.(\ref{eikave}) by setting $m=1$.
The $+$ label on the average indicates that the infrared singularities
are regulated in a quark-gluon plasma by the dynamically generated
masses of the initial  jet and  radiated gluon
as discussed in the previous section.
Note that this mechanism is  different from the regulation
of infrared singularities considered in Ref.\cite{gunion}
due to  form factors arising in collisions of color singlet hadrons.
In that case, interference amplitudes associated with radiation from
different hadronic constituents cancel both the $k_\perp=0$ and
the ${\bf k}_\perp ={\bf q}_\perp$ singularities. In a quark-gluon plasma
at the perturbative level the quasi-particles are color non-singlet
partons, and the singularities are regulated by medium polarization
effects\cite{selik3}.
Nevertheless, in both cases similar expressions arise
at the end.

With eqs.(\ref{rad},\ref{gunion1}),
we can define the ``radiation or color formation  factor'',
$ C_m(k)$ via
\begin{equation}
\frac{d^3 n_m}{d^3 k}\equiv C_m(k)\frac{d^3 n_1}{d^3 k}
\; \; . \label{defcm}\end{equation}
The result can be expressed as
\begin{equation}C_m(k)= \frac{1}{C_2^mC_Ad} \sum_{i=1}^m
\left(C_{ii} + 2Re \sum_{j=1}^{i-1}
C_{ij} e^{i(z_i-z_j)/\tau(k)}  F_{ij}(k)\right)
\; \; , \label{cm} \end{equation}
in terms of color coefficients,
\begin{equation}C_{ij}=Tr(a_m\cdots [c,a_i]\cdots a_1
a_1 \cdots [a_j,c] \cdots a_m)
\; \; , \label{clj} \end{equation}
and current correlation functions
\begin{equation}
F_{ij}(k)=\langle \vec{\jmath}_i \cdot\vec{\jmath}_{j} \rangle/
\langle |\vec{\jmath}_i|^2 \rangle
\; \; , \label{f2}\end{equation}
with $\vec{\jmath}_i$ given by eq.(\ref{jelem}).

\subsection{Color Coefficients}

The color coefficients in eq.(\ref{clj})
can be computed by repeated
 use of  basic $SU(N)$ relations for sums of products
of generators:
\begin{eqnarray}
aa&=&C_2{\bf 1}_d\; \; , \; \; [a,b]a=-\frac{C_A}{2}b\; \; ,
\nonumber \\
 aba&=&(C_2-C_A/2)b \; \; , \; \; [a,b] [b,a] = C_A C_2 {\bf 1}_d
\; \; . \label{sunrel} \end{eqnarray}
The diagonal coefficients  can  be seen to be identical to the
normalization factor in eq.(\ref{cm}):
\begin{eqnarray}
C_{ii}&=& Tr(C_2^{m-i}[c,a_i]C_2^{i-1}[a_i,c]) = C_2^m C_A d
\; \; .\label{cll} \end{eqnarray}
The off diagonal $j<i$ coefficients are, on the other hand,
\begin{eqnarray}
C_{ij}&=& Tr(C_2^{m-i}[c,a_i]a_{i-1}\cdots a_{j}  C_2^{j-1} [a_j,c]
a_{j+1}\cdots a_i)\nonumber \\
&=&-\frac{C_A^2}{4} C_2^{m-i+j-1}(C_2-\frac{C_A}{2})^{i-j-1} C_2 d\nonumber \\
&=&-{r_2} (1-{r_2})^{i-j-1} C_{ii}/2
\; \; , \label{clj2} \end{eqnarray}
where
\begin{equation}r_2=\frac{C_A}{2C_2} = \left\{ \begin{array}{ll}
N^2/(N^2-1) &{\rm for \; quarks \; with}\; C_2=C_F \\
1/2        &{\rm for \; gluons \; with}\; C_2=C_A
\end{array} \right.
\; \; .\label{r2} \end{equation}
The radiation formation factor is therefore
\begin{equation}C_m(k)= m- \frac{r_2}{1-r_2}
Re \sum_{i=1}^m \sum_{j=1}^{i-1}
(1-r_2)^{i-j}F_{ij}(k) e^{iL_{ij}/\tau(k)}
\; \; . \label{cm2} \end{equation}
For a single isolated scattering, of course,
\begin{equation}C_1(k)=1
\; \; . \label{c1} \end{equation}
For multiple scattering in the Bethe-Heitler limit,
corresponding to  $L_{ij}\gg \tau(k)$,
the phase factors average to zero, and the intensity of
induced radiation is simply additive in the number of scatterings,
i.e.,
\begin{equation}C_m(k) \approx m \; \; \; {\rm if} \; L_{ij} \gg \tau(k)
\; {\rm for\; all\;} i>j
\; \; . \label{cmlim1} \end{equation}
In the deep  LPM limit, where $\tau(k) \gg  L_{ij}$,
the destructive interference pattern summarized by the
 formation factor depends on the form of the current
correlation functions.
It is amusing to note
that  that the negative sign leading to destructive
interference in eq.(\ref{cm2}) arises in QCD from
the color algebra, eq.(\ref{clj2}),
 in contrast to QED where the destructive
pattern in eq.(\ref{qedj}) arises from
the opposite sign  of contributing momentum space amplitudes.

\subsection{Current Correlation Function}

In order to investigate the formal structure
of  the current correlation function, eq.(\ref{f2}),
we evaluate
\begin{eqnarray}
\langle  \vec{\jmath}_i \cdot\vec{\jmath}_{j} \rangle
&\propto&\left\langle \left(\frac{{\bf k}_\perp}{k_\perp^2} +
\frac{{\bf q}_{\perp i}-{\bf k}_\perp}{|{\bf q}_{\perp i}
-{\bf k}_\perp|^2}\right)\cdot
 \left(\frac{{\bf k}_\perp}{k_\perp^2} +
\frac{{\bf q}_{\perp j}-{\bf k}_\perp}{|{\bf q}_{\perp j}
-{\bf k}_\perp|^2}\right)
\right\rangle_+
\nonumber \\
&\approx& (1+H(k_\perp^2))^2/k_\perp^2
\; \; ,
 \label{j1j2}\end{eqnarray}
where in terms of the transverse vector, ${\bf J}\equiv(
{\bf q}_\perp-{\bf k}_\perp)/|{\bf q}_\perp-{\bf k}_\perp|^2$
\begin{eqnarray}
H(k_\perp^2)&=&\left\langle{\bf k}_\perp\cdot{\bf J}\right\rangle_+\equiv
{\bf k}_\perp\cdot\left\langle \frac{{\bf q}_\perp-{\bf k}_\perp}{
|{\bf q}_\perp-{\bf k}_\perp|^2}\right\rangle_+
\; \; . \label{hk}\end{eqnarray}
The approximate independence of the current correlations on
the indices, $i,j$, is only valid in a kinematic region
$k_{\perp} > xp_{\perp i}$
recalling eqs.(\ref{app}). In QED this restriction is severe
because the leading term,
$\epsilon_\perp \cdot {\bf k}_\perp/k_\perp^2$, from eq.(\ref{app}) cancels,
and the photon spectrum is peaked at $x\sim 1$.
In QCD, on the other hand, the radiated energy fraction, $xdn/dxd^2k_\perp$,
is approximately independent of $x$ from eq.(\ref{gunion1}).
Therefore, unlike in QED the regime
$x\ll 1$ is relevant in the case of QCD. We find below that the induced
radiation is indeed limited to $x< \lambda \mu^2/E_0 \ll 1$,
and thus the above approximation is justified.
This approximation cannot however be extended outside the
soft eikonal limit. For moderate $x< 1$ it clearly breaks
 down especially because
$p_{\perp i}^2$ grows approximately linearly with $i$
due to multiple scattering. For the general case,
the exact current element must be used
and the correlation function must be computed
from a  solution of a transport equation,
as first done by Migdal\cite{lpm}
for QED. We limit the discussion here to the soft eikonal regime.

The diagonal correlator, in the same limit is
proportional to the invariant gluon distribution
from a single collision:
\begin{eqnarray}
\langle  |\vec{\jmath}_i|^2 \rangle
&\propto&\left\langle \left|\frac{{\bf k}_\perp}{k_\perp^2} +
\frac{{\bf q}_\perp-{\bf k}_\perp}{|{\bf q}_\perp-{\bf k}_\perp|^2}
\right|^2
\right\rangle_+= \left\langle \frac{q_\perp^2}{k_\perp^2
|{\bf q}_\perp-{\bf k}_\perp|^2}\right\rangle_+
\nonumber \\
&=& (1+2 H(k_\perp^2)+H_2(k_\perp^2))/k_\perp^2
\; \; ,
 \label{j1j1}\end{eqnarray}
which involves a second function
\begin{eqnarray}
H_2(k_\perp^2)&=&k_\perp^2 \left\langle |{\bf J}|^2 \right\rangle_+=
\left\langle \frac{{k}_\perp^2}{
({\bf q}_\perp-{\bf k}_\perp)^2}\right\rangle_+
\; \; . \label{h2k}\end{eqnarray}
The approximate correlation function in the soft eikonal limit
is therefore independent of $i,j$ and given by
\begin{eqnarray}
F_{ij}(k)\approx F(k_\perp)=\frac{1+2H+H^2}{1+2H+H_2}
\; \; . \label{fijapp}\end{eqnarray}
Because $\langle |{\bf J}-\langle{\bf k}_\perp\cdot{\bf J}
\rangle/{k}_\perp^2  |^2\rangle \ge 0$, note that $H_2\ge H^2$
and consequently the current correlation function is bounded:
\begin{equation}
0\le F(k)\le 1
\; \; . \label{bound}\end{equation}
The upper bound is approached for ${k}_\perp^2\ll \langle
{q}_\perp^2 \rangle \sim \mu^2 $. In that soft region
 both $H\approx H_2\propto {k}_\perp^2/\langle {q}_\perp^2
\rangle\ll 1$, and
\begin{equation}
F_{ij}({k})\approx F(0)=1
\; \; . \label{f20lim}\end{equation}
The lower bound is approached in the opposite limit,
 ${k}_\perp^2\gg \langle {q}_\perp^2 \rangle$,
Formally, $H\approx -1 -\langle {q}_\perp^2 \rangle/{k}_\perp^2$
while $H_2\approx 1+ 3\langle {q}_\perp^2 \rangle/{k}_\perp^2$,
and consequently in that limit
\begin{equation}
F({k}) \propto \langle
{q}_\perp^2 \rangle/{k}_\perp^2
\; \; . \label{f2inflim}\end{equation}
The exact form interpolating between these limits
depends of course on the proper inclusion
of  polarization effects in the medium.

\subsection{The Factorization Limit}

For fixed $x\approx \omega/E_0\ll 1$
and  $k_\perp \rightarrow 0$, $F_{ij}\approx 1$, and
the formation length, $\tau(k)=xP^+/k_\perp^2$  becomes much longer
than the separation of the scattering centers. In this case,
the phase factors can be set to unity.
With the help of
\begin{equation}\sum_{i=1}^m \sum_{j=1}^{i-1}
(1-r_2)^{i-j}= \frac{1-r_2}{r_2}\left(m-\frac{1}{r_2}(1-(1-r_2)^m)\right)
\; \; , \label{sumij} \end{equation}
the radiation formation factor reduces to
\begin{eqnarray}
\lim_{k_\perp\rightarrow 0} C_m(k)&=&C_m^0
=\frac{1}{r_2} (1 -(1-r_2)^m) \nonumber \\
&=&\left( \begin{array}{ll}
1-1/N^2(1-1/(1-N^2)^{m-1})  &{\rm for \; quarks} \\
2(1-1/2^m)        &{\rm for \; gluons }
\end{array} \right.
\nonumber \\ &\;& \;\hspace{4in} .\label{cmlim2} \end{eqnarray}
Note also that for a given $r_2=C_A/2C_2$, $C_m^0$ approaches
$1/r_2$, independent of the number of collisions as $m\rightarrow\infty$.
This is the Factorization limit, in contrast to the
additive  Bethe-Heitler limit.
The saturation value  depends on the $SU(N)$ representation
of the jet parton and
causes the $C_A$ factor in eq.(\ref{gunion1}) to be replaced
by $2C_2$.
Interestingly, for quarks  in the fundamental representation
the destructive interference is so effective
that for $N=3$,  the final radiation intensity in the $k_\perp
\ll \langle q_\perp \rangle$ region after
many collisions is even slightly less, $1/r_2=8/9$,
than for a single isolated collision.
However, for incident gluon jets,
the induced radiation approaches twice that from a single
collision. For exotic hybrid partons in very high dimensional
representations
of $SU(N)$, the suppression effect in fact disappears altogether
for fixed $m$ as $C_A/mC_2\rightarrow 0$.
This dependence of the LPM effect on the representation
of the parton is a specific  non-abelian effect
in QCD.

\subsection{Ensemble Averaged Formation Factor}

For $0<k_\perp \ll \mu$, we can write
\begin{equation}C_m(\omega,{\bf k}_\perp)= C_m^0 + \frac{r_2}{1-r_2}
Re \sum_{i=1}^m \sum_{j=1}^{i-1}
(1-r_2)^{i-j} (1-F_{ij}(k)e^{i(z_i-z_j)/\tau(k)})
\; \; .\label{cm3} \end{equation}
To see analytically how $C_m$ interpolates between
$C_m^0$ and $m$ as a function
of the ratio of the mean free path to the formation
time, we average now
over the interaction $z_i$ points according to  linear kinetic theory.
Because  we restrict the discussion
here to the eikonal case, the  complication due to the full 3D
transport evolution can be neglected.
In linear kinetic theory the longitudinal
separation between successive scatterings,  $L_i=z_{i+1}-z_i$,
is  distributed simply as
\begin{equation}\exp(-L_i/\lambda)/\lambda
\; \; , \label{dtl} \end{equation}
where $\lambda=(\sigma_0 \rho v_0)^{-1}$ is the mean free path.
Therefore,
\begin{equation}\langle e^{i(z_i-z_j)/\tau(k)}
\rangle \approx  \left(\frac{1}{1-i\lambda/\tau(k)}\right)^{i-j}
\; \; . \label{ens1} \end{equation}
where $\tau(k)=2xE_0/k_\perp^2$.

The sum in eq.(\ref{cm2}) can then be performed
since in this soft eikonal limit
$F_{ij}(k)\approx F(k)\approx 1$
In that case eq.(\ref{sumij}) can be used
with the replacement
\begin{equation}1-r_2 \rightarrow
 \frac{1-r_2}{1-i\lambda/\tau(k)}
\; \; . \label{replace} \end{equation}
The resulting radiation formation factor reduces to
\begin{eqnarray}
C_m(k) &=&
m-\frac{m F(k)}{1+\chi^2(k)}
+ \frac{F(k)}{r_2}{\rm Re}\left(
\frac{(1-ir_2\chi(k))}{
(1-i\chi(k))^2}
\left[ 1-
\left(\frac{1-r_2}{1-ir_2\chi(k)}\right)^m \right] \right)
\; \; , \label{cf1} \end{eqnarray}
where the dimensionless function controlling the non-abelian LPM
effect is
\begin{equation}\chi(k)=\frac{\lambda}{\tau(k) r_2}
=\frac{\lambda k_\perp^2}{2 x r_2 E_0}
=\frac{C_2}{C_A}
\frac{\lambda k_\perp}{ \cosh(y)} \; \; . \label{r} \end{equation}
The last form is  in terms of the rapidity,
$y$ of the radiated gluon $(\omega=k_\perp \cosh y$).
Note that eq.(\ref{cf1}) satisfies  all the previous
limits considered ($m=1$, $k_\perp=0$,
$\lambda=\infty$).

For moderate large $m$ the term proportional to
 $(1-r_2)^m$ term can be neglected, and the  formation factor
simplifies to
\begin{eqnarray}
C_{m}(x,k_\perp)&\approx &
m\left(1-\frac{F(k)}{1+\chi^2(k)}\right)
+ \frac{F(k)}{r_2}\left(\frac{1-
(1-2r_2)\chi^2(k)}{(1+\chi^2(k))^2}\right)
 	\; \; . \label{cf2} \end{eqnarray}
This illustrates clearly how the radiation formation factor
interpolates between the $\tau=0$ and $\tau=\infty$
limits as a function of the dimensionless variable $\chi$.
However, it also shows that,
through the dependence on the current correlation function,
$F(k)$, the radiation formation factor  is actually a function
of two dimensionless variables, $\chi(k)=\lambda/r_2\tau$ and
 $k_\perp^2/\mu^2$. Thus, both the range, $\mu^{-1}$,
as well as the separation, $\lambda$, of the interactions
influences the final interference pattern.

For quarks, $r_2\approx 1$, while for gluons
$r_2=1/2$. In the collinear regime, $k_\perp \ll \mu$ regime,
$F(k)\approx 1$, and a further simplification occurs.
The radiation formation factors
for incident quarks and gluons
reduce then to the following simple ``pocket'' formulas

\begin{eqnarray}
C_m^q(k)&\approx& \frac{\tau^2(k)+m\lambda^2}{\tau^2(k)+\lambda^2}
\; ; , \nonumber \\
C_m^g(k)&\approx& \frac{4m\lambda^2}{\tau^2(k)+4\lambda^2}+
\frac{2}{(\tau^2(k)+4\lambda^2)^2}
\; \; .\label{cqg} \end{eqnarray}
While the above interpolation formula for quarks  can be
extrapolated  down to $m=1$, the gluon one only applies
for large $m$ because of
 the extra factor of two
 radiation in that case for large $m$.
We emphasize again the restriction  $x\mu \surd{m}\ll  k_\perp\ll \mu$
used in deriving the above expressions.

\section{Induced Soft Radiative Energy Loss}

An important  application of the radiation formation factor
is to the problem of calculation
of the radiative energy loss per unit length,
$dE/dz$, for a parton passing through
dense, color neutral matter with a mean free path, $\lambda\gg \mu^{-1}$.
We need only the incremental increase of the induced radiation
going from $m$ to $m+1$.
For moderately large $m$ and $k_\perp < \mu$
\begin{equation}dC_m/dm \approx \chi^2(k)/(1+\chi^2(k))
\; \; , \label{dcm} \end{equation}
with $\chi(k)$ given by eq.(\ref{r}).
Increasing $m\rightarrow m+1$,
the average increase of the interaction length is $\lambda$,
and thus from eq.(\ref{gunion1},\ref{defcm})
\begin{eqnarray}
dE_{soft}/dz&=&
\frac{d}{\lambda dm}\int dn_g \omega
\approx \int_0^\mu  dk_\perp^2 \int_{x_0}^{x_1} \frac{dx}{x}
\frac{\alpha_s C_A}{\pi \lambda k_\perp^2}
 \; \frac{xE_0\chi^2(x,k_\perp)}{1+
\chi^2(x,k_\perp)}
\; \; . \label{dedx1} \end{eqnarray}
The subscript ``soft'' is included
to emphasize  that the eikonal approximation,
used in deriving eq.(\ref{dcm}),
restricts its applicability to the
kinematic region $x\mu \surd m  < k_\perp < \mu$.
The limits on the fractional
energy loss are $x_0\approx k_\perp/E_0$
from kinematics, and $x_1\sim k_\perp/\mu $ from the
above restriction.
Note that we neglect the  $1/\surd m$ dependence
of $x_1$ since the Factorization limit is found
to be insensitive to this cutoff, while
the Bethe-Heitler limit must reduce in any case just the sum the radiation
from isolated ($m=1$) collisions.

The kinematic restriction  $k_\perp^2< \mu^2$,
also follows from   the destructive interference between
three gluon vertex and jet line amplitudes.
This point was already emphasized in \cite{gunion}
 but was missed in ref.\cite{ryskin},
where the domain of $k_\perp$ integration was allowed to extend
up to the kinematic limit, $k_\perp^2\le s/4=E_0 m/2$.
That lead to the erroneous conclusion that $dE/dz\propto E_0$,
in violation with the factorization theorems.

The  integral  over
$x$ can be performed by changing variables to $\chi$,
with the result
\begin{eqnarray}
\int_{x_0}^{x_1}
\frac{dx}{x} \frac{xE_0\chi^2}{1+\chi^2}
&=& \lambda_2 k_\perp^2 \left(\tan^{-1}(\lambda_2 k_\perp) - \tan^{-1}
(\lambda_2 k_\perp (\mu / E_0))\right)
\; \; , \label{dxint} \end{eqnarray}
where $\lambda_2\equiv \lambda/2r_2$.
We see that in the $E_0 \gg \lambda_2 \mu^2$
limit, the second term
is negligible. Furthermore,  since $\lambda \mu \sim 1/g \gg 1$ is
a basic assumption in our multiple
collision analysis, the first term  in the brackets
is approximately $\pi/2$.

The integral over $k_\perp$ is also analytic
and illustrates  how $dE/dz$ interpolates between
the Factorization and Bethe-Heitler limits.
We find that
\begin{equation}
dE_{soft}/dx \approx
{\textstyle \frac{1}{2}}\alpha_s
C_2\mu^2 \{L[ \lambda_2 \mu] - L[\lambda_2 \mu
(\mu / E_0)] \}
\; \; , \label{dedx2L} \end{equation}
where the interpolation function is
\begin{equation}
L[a]=\frac{2}{\pi}\left[ (1+\frac{1}{a^2})\tan^{-1}(a) - \frac{1}{a}
\right]
\; \; . \label{la} \end{equation}
Note that for $a\gg 1$,
$L[a]\approx 1-4/(\pi a)$. For $a\ll 1$,
on the other hand, $L[a]\approx 4x/3\pi$.
Since, $\lambda_2 \mu \gg 1$, the first term in the brackets
is always close to unity. However, the second term
depends on the dimensionless ratio, $\zeta\equiv\lambda \mu^2/ E_0$.
This ratio is large in the additive Bethe-Heitler limit and
small in the Factorization limit.

To see how eq.(\ref{dedx2L}) interpolates between
those two limits consider first  the approximate
 Factorization limit. We fix $\lambda_2 \mu\gg 1$
and send $E_0\rightarrow \infty$.
In that case, the second term in eq.(\ref{dedx2L}) can be neglected
and
\begin{equation}
dE_{soft}/dz \approx \frac{1}{2}\alpha_s
C_2\mu^2  \approx 2\pi \alpha_s^2 C_2 T^2
\; \; , \label{dedx2} \end{equation}
where
we used $\mu\sim gT$  to estimate the force range
in  a quark-gluon plasma at temperature $T$.
The result is thus sensitive to the square of the radiated
transverse momentum as suggested in \cite{highpt}.
Note that because we are only calculating the low
$k_\perp < \mu$ contribution, our derivation
 does  not allow us
to calculate logarithmic energy dependent factors as obtained
qualitatively in \cite{highpt}.
However, up to such logarithmic factors
eq.(\ref{dedx2}) demonstrates the approximate constant
behavior of the induced radiated energy loss
in the Factorization limit.

In the other extreme limit, we fix $E_0$
and send $\lambda \mu\rightarrow \infty$ so that
$\zeta\gg 1$. This is the dilute limit where
the mean free path exceeds the radiation formation length.
In this case the arguments of both terms
in eq.(\ref{dedx2L}) are large, and the small difference
leads to
\begin{equation}
dE_{soft}/dz \sim \frac{1}{2}\alpha_s
C_2 \mu^2 \frac{4 E_0}{\pi\mu^2 \lambda_2 }
\sim  \frac{E_0}{\lambda} \left(\frac{2\alpha_s C_A}{\pi}\right)
\; \; . \label{dedx3} \end{equation}
Note that in this limit we recover the linear dependence
of $dE/dz$ on the incident energy (modulo logarithms),
as in the Bethe-Heitler formula.

It is interesting to note
that in the additive regime the radiated energy loss
is proportional to $C_A$, as for a single scattering via eq.(\ref{gunion1}),
However, in the approximate Factorization
limit the  induced radiated energy loss
is proportional to the $C_2$ of the jet parton.
This means in practice
that gluons radiate $C_A/C_F=9/4$ more gluons than quarks
for $SU(3)$.
Recall, that the energy loss
due to elastic collisions for gluons
is also enhanced relative to quarks by  the same $C_A/C_F$
factor\cite{thoma}.
For comparison, the energy loss due to elastic collisions from
\cite{thoma}
is
\begin{equation}
dE_{el}/dz \approx \frac{4\pi}{3} C_2\alpha_s^2 T^2 \log(E_0 /\pi\alpha_s T)
\; \; . \label{dedxel} \end{equation}
Therefore, the total $dE/dx$ simply scales with $C_2$.
This scaling differs from the qualitative  estimates
in \cite{highpt} using  the single scattering
bremsstrahlung cross sections of \cite{gunion}.
It is interesting to note that both the elastic and radiated
energy loss is proportional to $\alpha_s^2$
and are comparable in magnitude up to uncertain logrithmic factors.

We emphasize that eq.(\ref{dedx2}) for the radiated energy loss
is only an order of magnitude
 estimate because we have not calculated  the contribution
from the non-factoring $k_\perp > \mu$ domain.
In order to improve the estimate,  all the
multiple  three gluon vertex
amplitudes added to those computed here.
That moderate high $k_\perp$ regime also requires
a more careful treatment of the current correlation function
as well as of the polarization mechanisms that regulate
infrared singularities.
A proper treatment of the above problems remains
an open theoretical challenge.

Finally, we comment on the comparison of eq.(\ref{dedx2})
to the bound on $dE/dx$ derived by Brodsky and Hoyer\cite{brod2}
based on the uncertainty principle.
For a radiated gluon carrying away a fraction $x$ of the incident
energy, $E_0$, with a given $k_\perp$,
the uncertainty in its formation length is $\tau(k)$
from eq.(\ref{tauk}). The induced energy loss for radiating
one gluon is therefore
bounded by
\begin{equation}
dE/dx < \langle x E/\tau(k) \rangle = \langle k^2_\perp \rangle/2
\sim \mu^2/2
\; \; . \label{brod}\end{equation}
Our estimate satisfies this bound because $\alpha_s \ll 1$
was assumed throughout our perturbative analysis.
In fact, we may interpret $C_2 \alpha_s$
roughly the probability of radiating one gluon
with $\tau(k)<\lambda$ between  multiple collisions.
That gluon is radiated in a cylindrical phase space
with approximate  uniform rapidity density
between $0<y<\log(\mu\lambda/r_2)$ and limited  $k_\perp < \mu$.

\section{Summary and Discussion}

In this paper, we initiated a  study of multiple collision
theory in pQCD concentrating on the
eikonal limit. We calculated elastic and
inelastic multiple collision amplitudes
for a high energy parton propagating though
a ``plasma'' of static target partons.
We showed that the assumption
of color neutrality was vital to recover the classical
parton cascade picture in both large and small
angle scattering. Our main simplifying
assumption was that  the mean free path ($\lambda\sim
1/g^2 T$) in the target
was large compared to the range of the interactions ($\mu^{-1}\sim
1/g T$).
In particular, we showed how the classical
Glauber scattering cross section eq.(\ref{eikonal}) emerges after
ensemble averaging.

The main focus of the paper was to derive eq.(\ref{sum2}),
which shows how the sum of the
induced gluon radiation amplitudes can be expressed
as a convolution of elastic multiple scattering
amplitudes and an effective color current, eq.(\ref{curr3}).
This result was derived for the soft radiation
($x\ll 1$ and $k_\perp \ll \mu$) regime.
The limitations of the approximations leading to that
result were also carefully analyzed.
We showed that the triple gluon diagrams can be neglected
in the soft limit but are important
to cut off the $k_\perp$ distributions on the scale $\mu$.
Only when $k_\perp\rightarrow 0$ can the effective
current be pulled out of the multiple collision integral,
and the radiation amplitude factored in momentum space
as in eq.(\ref{sum}). There is never a factorization
of amplitudes in color space.
However, the color neutrality condition, eq.(\ref{ensembl}),
greatly simplifies the   ensemble average of
the squared amplitude.  In addition,
the off diagonal contributions in momentum space
from $\vec{J}^c_{a_1\cdots a_m}(k;\{{\bf q}_{\perp i}\})\vec{J}^c_{a_1\cdots
a_m}(k;\{{\bf q}^\prime_{\perp i}\})$
drop out, as shown below eq.(\ref{mij410}),
if the transverse width of the target is large compared to the
interaction range. Under these conditions,
it was possible to calculate the induced radiation
spectrum from eq.(\ref{rad}).

We defined the radiation formation factor, eq.(\ref{cm}), as the ratio
of the induced radiation spectrum to the spectrum from an isolated
collision. That factor measures
the suppression of induced radiation
with formation length, $\tau(k)> \lambda$,
and reveals the non-abelian analog of the LPM effect.
 The novel role of the color
 algebra that leads to this destructive interference
pattern in QCD was shown in Eqs. (\ref{clj2},\ref{cm2}).
We showed how this  factor interpolates between
the saturated  Factorization limit and the additive
Bethe-Heitler limit. A compact ``pocket'' formula
for the formation factor was derived in eq.(\ref{cqg})
illustrating the essential features of that interpolation.

Finally, we applied the formation factor to estimate
the contribution of soft induced gluon radiation
to the energy loss per unit length, eq.(95).
The result
in the Factorization limit, eq.(\ref{dedx2}), was shown to be consistent with
the uncertainty principle bound of \cite{brod2}
with a numerical coefficient, $C_2 \alpha_s$,
that had a simple physical interpretation as the
number of induced gluons radiated in the limited
phase space  with rapidity
between zero and $\log(\mu\lambda/r_2)$ and
with $k_\perp<\mu$.
Up to un-calculated logarithmic factors the radiative
energy loss was found to be comparable to the elastic
energy loss\cite{thoma}.
We also showed how the linear energy dependence
of $dE/dz$ is recovered in the opposite (Bethe-Heitler) limit
when $\lambda\gg E/\mu^2$.

Naturally, many problems need further study.
Especially important will be
to extend the derivation of radiative cross sections
to the moderate $k_\perp > \mu$ regime to cover
the non-factorizable ``semi-hard'' regime for induced radiation.
We concentrated on the soft regime here to simplify
our task.
The ``semi-hard'' regime is however also
complicated by the necessity of having to consider in detail
the polarization effects that regulate ${\bf q}_\perp ={\bf k}_\perp$
singularities and also the necessity of computing
the current correlation functions  discussed in section 4.2.
The open theoretical question in this connection
 is to what extent, if any, can
a {\em classical} parton cascade transport model be constructed
that correctly simulates the many subtle interference
phenomena of pQCD in the multiple collision domain.\\[3ex]

Acknowledgements: Stimulating
discussions with S. Brodsky, A. Mueller, B. Muller, M. Grabiak,
Sovand P. Danielewicz
are gratefully acknowledged. This work was initiated
while M.G. was at LBL and X.N. Wang was at Duke University.\\[2ex]

\section*{Appendix A: Large Angle Elastic Cascade Limit}

We consider here the elastic coincidence scattering
in the special idealized case that
the coordinates  ${\bf x}_i $ are fixed and the energy
is large to see how the classical billiard ball
formula emerges. We rewrite  the phases in eq.(\ref{mij40}) as
\begin{eqnarray}
{\bf p}_k\cdot{\bf R}_k
&\approx& {\bf P}_k\cdot {\bf R}_k
- \frac{L_k}{2P_0}({\bf p}_{\perp k} - {\bf P}_{\perp k})^2\; \; , \label{ph2}
\end{eqnarray}
with ${\bf P}_k=P_0{\bf R}_k/{R}_k$ and
\begin{eqnarray}
{\bf P}_{\perp k}&=& P_0 {\bf r}_{\perp k}/R_k\approx P_0 {\bf r}_{\perp k}/L_k
\; \; . \label{kinemij} \end{eqnarray}
Note  that ${\bf p}_{\perp k}
\cdot{\bf r}_{\perp k}=2{\bf p}_{\perp k}
\cdot{\bf P}_{\perp k}(L_k/2P_0)$ and
$P_0 L_k
\approx {\bf P}_k\cdot {\bf R}_k - {\bf P}_{\perp k}^2(L_k/2P_0)$.
Substituting eq.(\ref{ph2}) into eq.(\ref{mij40}) and shifting the
${\bf p}_{\perp k}$ integration, we find that
\begin{eqnarray}
I_{ji}(p_j,p_{i-1})&\approx& e^{+i\phi_{ji}}
\int \left\{ \prod_{k=i}^{j-1}
\frac{d^2{\bf p}_{\perp k}}{(2\pi)^2}
 \frac{ e^{-i\frac{L_k}{2P_k}{\bf p}_{\perp k}^2} }{
2P_0}\right\}
A_j({\bf Q}_j-{\bf p}_{\perp j-1})\nonumber \\
&\; & \hspace{0.5in} \times \left\{\prod_{l=i}^{j-1}
 A_l({\bf Q}_l+{\bf p}_{\perp l}-{\bf p}_{\perp (l-1)}) \right\}
\; \; ,  \label{mij5} \end{eqnarray}
where the intermediate  classical momentum transfers, denoted by
\begin{equation}{\bf Q}_k= {\bf P}_k-{\bf P}_{k-1} \; \; \;
{\rm for} \; \; i<k<j
\; \; , \label{vqk} \end{equation}
have dominantly  transverse components.
The endpoint momentum transfers are given
${\bf Q}_j={\bf p}_j-{\bf P}_{j-1}$ and ${\bf Q}_{i}={\bf P}_i-{\bf p}_{i-1}$,
and the  external phase is
\begin{equation}\phi_{ji}=\sum_{l=i}^{j-1}{\bf P}_l\cdot{\bf
R}_l=\sum_{l=i}^{j}{\bf Q}_l\cdot{\bf x}_l+
i{\bf p}_j\cdot{\bf x}_j-i{\bf p}_{i-1}\cdot{\bf x}_{i-1}
\; \; . \label{phiij} \end{equation}
A simplification occurs in the high energy when
the ${\bf x}_i$ are fixed
because the $Q_{\perp k}$  increase  linearly with $E_0$,
 while the transverse momentum
integrals are limited by the oscillating phase factors to
$p_{\perp k} {\mathrel{\lower.9ex\hbox{$\stackrel{\displaystyle <}{\sim}$}}}
(P_0/L_k)^{1/2}$.
Hence $Q_{\perp k}\gg p_{\perp k}$ for high enough energies,
and we can expand the potentials
around ${\bf Q}_{\perp k}$. This is equivalent to the stationary
phase approximation.
The region of validity of that approximation
can be clarified by writing
\begin{equation} {\bf Q}_{\perp k}=
P_0 ({\bf r}_{\perp k}/L_k -{\bf r}_{\perp k-1}/L_{k-1})
\equiv
P_0 \Delta {\bf r}_{\perp k}/L_k
\; \; . \label{vqk3} \end{equation}
The condition that ${Q}_{\perp k}\gg (P_0/L_k)^{1/2}$
is thus equivalent to requiring that the transverse momentum transfer be
large enough to resolve the transverse separation of the scattering centers:
\begin{equation}{\bf Q}_{\perp k}\cdot \Delta {\bf r}_{\perp k} \gg \hbar
\; \; . \label{cond2} \end{equation}
For {\em fixed} ${\bf r}_{\perp k}$ this condition is always satisfied
for sufficiently large energies, and
thus the stationary phase integrals
can then be evaluated using
\begin{equation}\int \frac{d^2{\bf p}_\perp}{2\pi i P} e^{-ip_\perp^2 L/2P}
= \frac{1}{L}
\; \; , \label{station} \end{equation}
In this limit $M_{ji}$ reduces to the simple factorized
form
\begin{eqnarray}
M_{ji}(p_j,p_{i-1}) &\approx &\delta(ji) e^{+i{\bf Q}_j\cdot{\bf x}_j}
A_j({\bf Q}_j)
 \prod_{k=i}^{j-1}\left\{ e^{i\pi/2}
 e^{+i{\bf Q}_k\cdot{\bf x}_k}
A_k({\bf Q}_k)/(4\pi R_k) \right\}
\; \; . \label{mij5c} \end{eqnarray}
Note again that in the non-Abelian case the matrix ordering
from $j$ to $i$ is essential.

After squaring and integrating over the magnitude of the
final momentum, averaging over initial colors and summing over
final, the above factorized form leads via eq.(\ref{dsido})
to the classical billiard ball
formula
\begin{equation}d\sigma_{ji}/d\Omega = d\sigma_{i}/d\Omega_i\left\{
\prod_{k=i+1}^{j} d\sigma_{k}/R_k^2 d\Omega_k \right\}
\; \; . \label{cascade} \end{equation}
We emphasize that color neutrality of the medium and
large transverse momentum transfers
are  essential to recover this simple cascade limit
in which  the direction of all intermediate momenta are fixed
by geometry.



\section*{Appendix B: Derivation of Eq.(39)}

Technical details of the derivation of the radiation amplitude
from intermediate jet lines are given here.
The integration over $p_{zj}$
in eq.(\ref{mrim0})
can be
expressed as a sum of two terms using eq.(\ref{del}).
 The term containing  the $\Delta(p_j-k)$
propagator leads to
\begin{eqnarray}
M_{cj}^+  &=& ig\tilde{\delta}(m1) e^{-i{\bf p}_m\cdot{\bf x}_m}e^{+i{\bf
p}_0\cdot {\bf x}_1}
\int \frac{d^3p_j}{(2\pi)^3}
\frac{e^{+i{\bf p}_j\cdot{\bf R}_j}
e^{-i{\bf k}\cdot{\bf x}_{j+1}}}{{P}_\omega^2
- ({\bf p}_j-{\bf k})^2 +i\epsilon}
\nonumber \\
&\; & \hspace{0.5in} \times
\frac{\epsilon(p_j-k)}{
k(p_j-k)}
I_{m,j+1}(p_m,p_{j}-k) T_c I_{j,1}(p_j,p_0)
\; \; . \label{mrip} \end{eqnarray}
The term containing  the $\Delta(p_j)$
propagator leads to
\begin{eqnarray}
M_{cj}^-  &=& -ig\tilde{\delta}(m1) e^{-i{\bf p}_m\cdot{\bf x}_m}e^{+i{\bf
p}_0\cdot {\bf x}_1}
\int \frac{d^3p_j}{(2\pi)^3}
\frac{e^{+i{\bf p}_j\cdot{\bf R}_j}e^{-i{\bf k}\cdot{\bf x}_{j+1}}}{P_0^2 -
{\bf p}_j^2 +i\epsilon}
\nonumber \\
&\; & \hspace{0.5in} \times
\frac{\epsilon p_j}{
k p_j}
I_{m,j+1}(p_m,p_{j}-k) T_c I_{j,1}(p_j,p_0)
\; \; . \label{mrim} \end{eqnarray}
In $M_{cj}^+$ the contour integral over $p_{zj}$ sets
$p_j -k$ on shell with
\begin{eqnarray}
p_{zj}&\approx& k_z + {P}_\omega - ({\bf p}_{\perp j} -{\bf
k}_\perp)^2/2{P}_\omega
\nonumber \\
&\approx &P_0 - ({\bf p}_{\perp j} -{\bf k}_\perp)^2/2P_0 + (k_z-\omega/v_0)
\; \; , \label{pzjp} \end{eqnarray}
 while in $M_{cj}^-$ it sets $p_j$ on shell with $p_{zj}
\approx P_0 - {\bf p}_{\perp j}^2/2P_0$.
In both cases the residue of the propagator can be
well approximated by $1/2P_0$ in the high energy limit when $x\ll 1$.
 However, it is essential to keep track of the difference,
$P_0-P_\omega=\omega/v_0$ in computing the phases.
The phase in the integrand of  $M_{cj}^+$ is given by
\begin{eqnarray}
{\bf p}_j\cdot{\bf R}_j-{\bf k}\cdot{\bf x}_{j+1}&=&
({\bf p}_j-{\bf k}){\bf R}_j-{\bf k}\cdot{\bf x}_{j}
\nonumber \\
&\approx& {\bf P}_j\cdot {\bf R}_j
-\frac{L_j}{2P_0}({\bf p}_{\perp j} -{\bf k}_\perp - {\bf P}_{\perp j})^2-
\omega R_j/v_0 -{\bf k}\cdot{\bf x}_{j}
\; \; . \label{phsp1} \end{eqnarray}
Note that the replacement $\omega L_j/v_j\rightarrow \omega R_j/v_j$
above is valid  either when $r_{\perp j}\sim d \ll L_j$
or  $P_0\rightarrow \infty$.
The phase in  $M_{cj}^-$ is, on the other hand,
\begin{equation}{\bf p}_j\cdot{\bf R}_j-{\bf k}\cdot{\bf x}_{j+1}
\approx {\bf P}_j\cdot {\bf R}_j
-\frac{L_j}{2P_0}({\bf p}_{\perp j} - {\bf P}_{\perp j})^2
-{\bf k}\cdot{\bf x}_{j+1}
\; \; . \label{phsp} \end{equation}
Shifting the ${\bf p}_{\perp j}$ integration
to ${\bf p}_{\perp j}+{\bf k}_\perp+{\bf P}_{\perp j}$ and
${\bf p}_{\perp j}+{\bf P}_{\perp j}$ in $M_{cj}^\pm$ respectively,
\begin{eqnarray}
M_{cj}^+  &\approx& g\tilde{\delta}(m1)
e^{-i{\bf p}_m\cdot{\bf x}_m}e^{+i{\bf p}_0\cdot {\bf x}_1}
\int \frac{d^2{\bf p}_{\perp j}}{(2\pi)^2}
\frac{1}{2P_0}
e^{i{\bf P}_j\cdot{\bf R}_j}e^{-i\frac{L_j}{2P_0}{\bf p}_{\perp j}^2}
\nonumber \\
&\; & \hspace{0.2in} \times
\left(
\frac{\epsilon \tilde{p}_j}{
k\tilde{p}_j} e^{-i(\omega R_j/v_j+{\bf k}\cdot{\bf x}_{j})} \right)
I_{m,j+1}(p_m,\tilde{p}_j) T_c I_{j,1}(p_j,p_0)
\; \; , \label{mjp2} \end{eqnarray}
\begin{eqnarray}
M_{cj}^-  &\approx& -g\tilde{\delta}(m1)
e^{-i{\bf p}_m\cdot{\bf x}_m}e^{+i{\bf p}_0\cdot {\bf x}_1}
\int \frac{d^2{\bf p}_{\perp j}}{(2\pi)^2}
\frac{1}{2P_0}
e^{i{\bf P}_j\cdot{\bf R}_j}e^{-i\frac{L_j}{2P_0}{\bf p}_{\perp j}^2}
\nonumber \\
&\; & \hspace{0.2in} \times
\left(
\frac{\epsilon p_j}{k
p_j}  e^{-i{\bf k}\cdot{\bf x}_{j+1}} \right)
I_{m,j+1}(p_m,p_{j}-k) T_c I_{j,1}(p_j,p_0)
\; \; . \label{mrip2} \end{eqnarray}
Note that in $M_{cj}^-$, the on-shell $p_j$ and $\tilde{p}_j$
are given by Eqs.(\ref{pjmu1},\ref{pjmu2}).
Therefore both the off-shell amplitude
$I_{m,j+1}(p_m,p_{j}-k)$ and the on-shell amplitude
$I_{m,j+1}(p_m, \tilde{p}_j)$ in the integrands
above are evaluated with the same
shifted incident energy, $E_0 - \omega$,
and approximately the same shifted incident longitudinal momentum,
${P}_\omega \approx P_0 - \omega/v_0$.

The energy shift of $I_{m,j+1}$ above leads in the high energy limit
 to a phase shift of those amplitudes
relative to the case without radiation.
To see this, note from eq.(\ref{mij3}) that
\begin{equation}I_{m,j+1}(p_m,p_{j}-k)=
\int \left\{ \prod_{k=j+1}^{m-1}
\frac{d^3{\bf p}_k}{(2\pi)^3}
 \frac{e^{i\pi/2}e^{+i{\bf p}_k \cdot{\bf R}_k}}{
P_\omega^2-{\bf p}_k^2+i\epsilon}\right\}
A_m({\bf p}_m-{\bf p}_{m-1})  \cdots A_{j+1}({\bf p}_{j+1}-{\bf p}_j+{\bf k})
\; \; . \label{mij32} \end{equation}
The contour integrals over the $p_{zk}$ then fix
the phases to be
\begin{eqnarray}
{\bf p}_k\cdot{\bf R}_k &\approx& {P}_\omega L_k - p_{\perp
k}^2(L_k/2{P}_\omega)
+ {\bf p}_{\perp k}
\cdot{\bf r}_{\perp k} \nonumber\\
&\approx& {\bf P}_k\cdot {\bf R}_k
- \frac{L_k}{2P_0}({\bf p}_{\perp k} - {\bf P}_{\perp k})^2 -\omega R_k/v_0
\; \; . \label{ph12} \end{eqnarray}
Therefore, there is an additional phase shift $\omega L_k/v_0\approx
\omega R_k/v_0$
for each intermediate line.
This phase shift has a simple physical interpretation.
Noting that the classical transit time
between centers at ${\bf x}_k$ and ${\bf x}_{k+1}$
is
\begin{equation}\Delta t_k \approx R_k/v_0
\; \; , \label{trans} \end{equation}
the additional phase shift is due to the time delay of a wavefront
propagating with a frequency  reduced by $\omega$.
Defining the classical interaction time at center, $k$, as
\begin{equation}t_k= \sum_{i=1}^{k-1} \Delta t_i
\; \; , \label{tk} \end{equation}
with $t_1\equiv0$, the net extra phase can be  factored out
from
eq.(\ref{mij32}) as
\begin{eqnarray}
I_{m,j+1}(p_m,p_j-k)&\approx& e^{-i\omega(t_m-t_{j+1})}
e^{+i\phi_{m,j+1}}
\int \left\{ \prod_{k=j+1}^{m-1}
\frac{d^2{\bf p}_{\perp k}}{(2\pi)^2}
 \frac{ e^{-i\frac{L_k}{2P_0}({\bf p}_{\perp k} - {\bf P}_{\perp k})^2} }{
2P_0}\right\} \nonumber \\
&\; & \hspace{0.5in} \times
A_m({\bf p}_m-{\bf p}_{m-1})  \cdots A_{j+1}({\bf p}_{j+1}-{\bf p}_j+{\bf k})
\; \; , \label{mij4tild} \end{eqnarray}
where $\phi_{m,j+1}$ is the phase without
radiation given via eq.(\ref{phiij}).
In eq.(\ref{mij4tild}) we again
used the condition $x\ll 1$ in replacing
the residue $1/{P}_\omega$ by  $1/P_0$.

Note that the longitudinal momentum transfer
in all the potential is still small
since all the intermediate longitudinal momenta are shifted
by approximately $k_z$. If in addition to $x\ll 1$,
 we consider radiation with  $k_\perp\ll \mu$,
then the arguments of all the potentials
inside the integrand can be  approximated by those
in eq.(\ref{mij40}). Only the extra $\omega$ dependent
phase must be kept.
Therefore, in this soft limit
\begin{equation}I_{m,j+1}(p_m,p_j-k)\approx e^{-i\omega(t_m-t_{j+1})}
I_{m,j+1}(p_m,p_j)
\; \; , \label{tilij1} \end{equation}
where the right hand side is to be evaluated ignoring the soft
radiation via eq.(\ref{mij40}).
Similarly, it follows that for soft radiation
\begin{equation}I_{m,j+1}(p_m,\tilde{p}_j)\approx  e^{-i\omega(t_m-t_{j+1})}
I_{m,j+1}(p_m,p_j)
\; \; , \label{tilij2} \end{equation}
involving the same phase shift as in eq.(\ref{tilij1}).
Combining these results we obtain eq.(39).


\section*{Appendix C: Derivation of Eq.(56)}

The details of the derivation of the radiation amplitudes
involving one three gluon vertex are given here.
{}From eq.(\ref{mij41})
\begin{eqnarray}
M_{j-1,1}(p_{j-1},p_0)&=& 2\pi \delta(p_{j-1}^0-E^0)(a_{j-1}\cdots a_{1})
(-ig)^{j-1} (2E_0) e^{-i(p_{z(j-1)}-P_0)z_{j-1}}
\nonumber \\
 &\;& \; \; \; \times \int \prod_{k=1}^{j-1} \left(
\frac{d^2 q_{\perp k}}{(2\pi)^2}  e^{-i{\bf q}_{\perp k}\cdot {\bf x}_{\perp
k}}
V^{a_k}_k({\bf q}_{\perp k}) \right) (2\pi)^2
\delta({\bf p}_{\perp (j-1)}  - \sum_{k=1}^{j-1} {\bf q}_{\perp k})
\nonumber \\
&\; & \hspace{2in} . \label{eik1} \end{eqnarray}
\begin{eqnarray}
M_{m,j+1}(p_m,p_j)&=& 2\pi \delta(p_m^0-p_j^0)(a_m\cdots a_{j+1})
(-ig)^{m-j} (2E_0) e^{-i(p_{zm}-P_\omega)z_m} e^{-i(P_\omega-p_{zj})z_{j+1}}
\nonumber \\
 &\;& \times \int \prod_{k=j+1}^m \left(
\frac{d^2 q_{\perp k}}{(2\pi)^2}  e^{-i{\bf q}_{\perp k}\cdot
{\bf x}_{\perp k}}
V^{a_k}_k({\bf q}_{\perp k}) \right) (2\pi)^2
\delta({\bf p}_{\perp m} - {\bf p}_{\perp j} - \sum_{k=j+1}^m
{\bf q}_{\perp k})
\nonumber \\
&\; & \hspace{2in} . \label{eik2} \end{eqnarray}
We have utilized above that $p_{z0}=P_0\approx E_0$ and that
the energy of internal lines $k<j$ is $E_0$, while for
$k\ge j$ it is $E_0-\omega$ with $p_{zk}=P_\omega=P_0-\omega/v_0$
on those lines.
Note that the amplitude $M_{m,j+1}$ differs from the case of no radiation
eq.(\ref{mij41}) by only a $z$ dependent  phase  obtained by replacing
$P_0$ with $P_\omega$ in eq.(\ref{phasz}).
In the eikonal limit eq.(\ref{mcj}) can be expressed as
\begin{eqnarray}
G_{j1}^c &=& 2\pi \delta(p_j^0-p_{j-1}^0+\omega)[c,a_j]
(-ig) (2E_0) e^{-i(p_{zj}-p_{z(j-1)}+k_z)z_{j}}
\nonumber \\
 &\;& \times \int
\frac{d^2 q_{\perp j}}{(2\pi)^2}  e^{-i{\bf q}_{\perp j}\cdot
{\bf x}_{\perp j}}
V^{a_j}_j({\bf q}_{\perp j})  (2\pi)^2
\delta({\bf q}_{\perp j}- ({\bf p}_{\perp j} -
{\bf p}_{\perp (j-1)}+{\bf k}_\perp))
\nonumber \\
 &\;& \times \left[\frac{-i\Gamma(k,p_j,p_{j-1})}{\omega^2
-(p_{zj}- p_{z(j-1)})^2-({\bf q}_{\perp j} -{\bf k}_\perp)^2} \right]
\; \;  , \label{eik3} \end{eqnarray}
where we used
$-iT_b f_{ba_jc}=[c,a_j]$ and note that the vertex function  is
\begin{eqnarray}
\Gamma(k,p_j,p_{j-1})
&=&4E_0 (p_j-p_{j-1})\epsilon(k)+4\omega p_j\epsilon(k)
\nonumber \\
&\;& \hspace{0.2in} -(p_j+p_{j-1})(p_j-p_{j-1}+2k)\epsilon^0(k)
\; \; . \label{gamma1} \end{eqnarray}

Inserting these expressions into eq.(\ref{gjm}),
the integrals over
 the $p_{j-1}^0,p_j^0$
variables give rise again to an overall $\tilde\delta(m1)$ factor
eq.(\ref{delm}) and set $p_{j-1}^0=E_0$ and $p_j^0=E_\omega=E_0-\omega$.
 Then as in eq.(\ref{mrip}) we integrate the contour over
$p_{zj}$ and $p_{z(j-1)}$ keeping only the residues at the poles
of the propagators since  $L_j\gg d$.
Because of the  three propagators involved,
$\Delta(p_j)\Delta(p_j-p_{j-1})\Delta(p_{j-1})$,
those integrals give rise to three  contributions corresponding
to forward scattering with  two of the three internal lines set on shell.
A fourth contribution corresponding to backscattering of the
gluon near $z_j$ is suppressed in the
kinematic range $E_0\gg\omega\gg max(k_\perp, q_{\perp j})$
of interest here.  The kinematic variables of the three forward amplitudes
after the contour integration
 are then as follows:
In case 1,  $p_j^2=p_{j-1}^2=0$, $(p_j-p_{j-1})^2
\approx -({\bf q}_{\perp j}-{\bf k}_\perp)^2$,
 with ${\bf p}_{\perp j}-{\bf p}_{\perp(j-1)}
={\bf q}_{\perp j}-{\bf k}_\perp$, and
\begin{eqnarray}
p_{j-1}&=& (E_0,E_0-p_{\perp(j-1)}^2/2E_0,{\bf p}_{\perp(j-1)})
 \nonumber \\
p_{j}&= & (E_0-\omega,E_0-\omega-p_{\perp j}^2/2E_0,{\bf p}_{\perp j})
\; \; . \label{kin1} \end{eqnarray}
In case 2, $(p_j-p_{j-1})^2=p_{j-1}^2=0$,
$p_j^2\approx -({\bf q}_{\perp j}-{\bf k}_\perp)^2/x$, and
\begin{eqnarray}
p_{j-1}&=& (E_0,E_0-p_{\perp(j-1)}^2/2E_0,{\bf p}_{\perp(j-1)})
 \nonumber \\
p_{j}&=& (E_0-\omega,E_0-\omega+({\bf q}_{\perp j}-
{\bf k}_\perp)^2/2\omega,{\bf p}_{\perp j})
\; \; . \label{kin2} \end{eqnarray}
Finally, in case 3 $(p_j-p_{j-1})^2=p_j^2=0$,  $p_{j-1}^2\approx
({\bf q}_{\perp j}-{\bf k}_\perp)^2/x$, and
\begin{eqnarray}
p_{j-1}&=& (E_0,E_0-({\bf q}_{\perp j}-{\bf k}_\perp)^2/2\omega,{\bf
p}_{\perp(j-1)})
 \nonumber \\
p_{j}&=& (E_0-\omega,E_0-\omega-p_{\perp j}^2/2E_0,{\bf p}_{\perp j})
\; \; . \label{kin3} \end{eqnarray}
We assume that $\omega^2 \gg
({\bf q}_{\perp j}- {\bf k}_\perp )^2$.
It is remarkable that the residue of the product of propagators
is approximately same up to a sign in all three case with
\begin{equation}(-1)Res( \Delta(p_j)\Delta(p_j-p_{j-1})\Delta(p_{j-1})) \approx
\pm
\frac{1}{({\bf q}_{\perp j}- {\bf k}_\perp )^2(2E_0)^2}
\; \; , \label{res3} \end{equation}
with $+$ for cases 1 and 2 and $-$ for case 3.
Also the vertex factor turns out to be approximately
the same in all three cases
\begin{equation}\Gamma
\approx -4E_0\vec{\epsilon}_\perp\cdot({\bf q}_{\perp j}- {\bf k}_\perp)
\nonumber \\
\; \; . \label{gamma2} \end{equation}
For a jet parton with mass, $\mu$, the singularity
in eq.(\ref{res3}) at ${\bf q}_{\perp j}={\bf k}_\perp$
is automatically regulated as in eq.(\ref{app}) by
\begin{equation}
|{\bf q}_{\perp j}-{\bf k}_\perp|^{-2} \rightarrow (|{\bf q}_{\perp j}-{\bf
k}_\perp|^{2}+ x^2\mu^2)^{-1}
\; \; . \label{app2} \end{equation}
In higher order, the inclusion of the
gluon proper self energy tensor in the medium would
replace this infrared regulator by the effective dynamic mass of the gluons.

Only the $z$ dependent phase,
\begin{equation}\phi_z=
-(P_\omega-p_{zj})z_{j+1}-(p_{zj}-p_{z(j-1)}+k_z)z_j-(p_{z(j-1)}-P_0)z_{j-1}
\; \; , \label{phsg} \end{equation}
is found  to be case dependent.
Evaluating with eq.(\ref{kin1},\ref{kin2},\ref{kin3}), we find that
\begin{eqnarray}
\phi=(\omega-k_z)z_j + \left\{
\begin{array}{ll}
0 \; \; \; & {\rm case } \; 1\; \\
L_j/\tau_j\; \; \;  &
{\rm case } \; 2\; \\
-L_{j-1}/\tau_j \; \; \; &
{\rm case }\; 3
\end{array}  \right. \nonumber \\
\; \; , \label{phi2} \end{eqnarray}
where $\tau_j\equiv
2\omega/({\bf q}_{\perp j}-{\bf k}_\perp)^2$ is the lifetime
of the virtual jet line in cases 2 and 3, and $L_j=z_{j+1}-z_j$
is the longitudinal distance between adjacent scattering centers.
Together with the relative signs in eq.(\ref{res3}), the phase factors
in the three cases sum to
\begin{equation}e^{i(\omega-k_z)z_j}(1+e^{iL_j/\tau_j}-e^{-iL_{j-1}/\tau_j})
\equiv e^{iz_j/\tau(k)}f_j(k)
\; \; . \label{sumphas} \end{equation}
We therefore find a new  interference form
factor, $ f_j$ , that involves the
separation distances between adjacent centers and
the lifetime of the virtual jet state, $\tau_j$.
For well separated centers, in the sense that
the mean free path $\lambda =
\langle (z_{i+1}-z_{i}) \rangle \gg
\tau_j$,
the extra phases in $f_j$ average to zero and $f_j \approx 1$.
Also in
the extreme opposite limit, $\lambda \ll
\tau_j$,  extra terms tend to cancel again leading
to $f_j \approx 1$.
In particular, for ${\bf k}_\perp ={\bf q}_{\perp j}$, $f_j = 1$.
As a rough form illustrating these limits,
\begin{equation}f_j(k)\sim
 1+ 2i\sin(\lambda({\bf q}_{\perp j}-{\bf k}_\perp)^2
/2\omega)
\; \; .\label{fj} \end{equation}
In the general case, $f_j\ne 1$ reflects the effects of final state
cascading of the emitted gluon.
Another  important limit is
$q_j=0$ or $k_\perp \gg q_{\perp j}$.
In this limit, $\tau_j\approx\tau(k)=1/(\omega-k_z)$, and $f_j$
reduces to
\begin{equation}
f_j\approx 1+e^{iL_{j}/\tau(k)}-e^{iL_{j-1}/\tau(k)}
\; \; . \label{q0lim}\end{equation}
For radiation with formation time much less {\em or} much
longer than the mean free path, $f_j\approx 1$.
Therefore, except in the restricted kinematic domain
where $\tau_j$ or $\tau(k)$ is on the order of the mean free
path, this extra interference effect can be neglected, and
$f_j$ can be set to unity.
Combining these results, we obtain eq.(56).

\end{document}